\documentclass[trackchanges,twocolumn]{aastex701}
\usepackage{amsmath}

\newcommand{\EP}{\textit{Einstein Probe}}
\newcommand{\ep}{\textit{EP}}
\newcommand{\mephisto}{\textit{Mephisto}}
\newcommand{\fermi}{\textit{Fermi}}
\newcommand{\svom}{\textit{SVOM}}
\newcommand{\swift}{\textit{Swift}}

\begin{document}

\title{A Shell Collision Scenario for EP251222b/GRB 251222A: Connecting Prompt Emission and Double-Peaked Afterglow}

\author[0009-0001-9846-545X]{Run Shi}
\affiliation{Purple Mountain Observatory, Chinese Academy of Sciences, Nanjing 210008, China}
\affiliation{School of Astronomy and Space Sciences, University of Science and Technology of China, Hefei 230026, China}
\email[]{runshi@pmo.ac.cn}

\author[0000-0002-4304-2759]{Di Xiao}
\affiliation{Purple Mountain Observatory, Chinese Academy of Sciences, Nanjing 210008, China}
\affiliation{School of Astronomy and Space Sciences, University of Science and Technology of China, Hefei 230026, China}
\affiliation{State Key Laboratory of Radio Astronomy and Technology, Purple Mountain Observatory, Chinese Academy of Sciences, 10 Yuanhua Road, Nanjing 210023, People’s Republic of China}
\email[show]{dxiao@pmo.ac.cn}

\author[0000-0001-7225-2475]{Brajesh Kumar}
\affiliation{South-Western Institute for Astronomy Research, Yunnan University, Kunming, Yunnan 650500, People's Republic of China}
\affiliation{Yunnan Key Laboratory of Survey Science, Yunnan University, Kunming, Yunnan 650500, People's Republic of China}
\email[]{brajesh@ynu.edu.cn}

\author[0009-0005-0170-192X]{Yifang Liang}
\affiliation{Purple Mountain Observatory, Chinese Academy of Sciences, Nanjing 210008, China}
\affiliation{School of Astronomy and Space Sciences, University of Science and Technology of China, Hefei 230026, China}
\email[]{yfliang@pmo.ac.cn}

\author[0000-0002-8109-7152]{Guowang Du}
\affiliation{South-Western Institute for Astronomy Research, Yunnan University, Kunming, Yunnan 650500, People's Republic of China}
\affiliation{Yunnan Key Laboratory of Survey Science, Yunnan University, Kunming, Yunnan 650500, People's Republic of China}
\email[]{dugking@ynu.edu.cn}

\author[0000-0001-9648-7295]{Jinjun Geng}
\affiliation{Purple Mountain Observatory, Chinese Academy of Sciences, Nanjing 210008, China}
\affiliation{State Key Laboratory of Radio Astronomy and Technology, Purple Mountain Observatory, Chinese Academy of Sciences, 10 Yuanhua Road, Nanjing 210023, People’s Republic of China}
\email[show]{jjgeng@pmo.ac.cn}

\author[0009-0008-6247-0645]{Wenlong Zhang}
\affiliation{Purple Mountain Observatory, Chinese Academy of Sciences, Nanjing 210008, China}
\affiliation{School of Astronomy and Space Sciences, University of Science and Technology of China, Hefei 230026, China}
\email[]{wlzhang@pmo.ac.cn}

\author[0009-0006-1010-1325]{Yuan Fang}
\affiliation{South-Western Institute for Astronomy Research, Yunnan University, Kunming, Yunnan 650500, People's Republic of China}
\affiliation{Yunnan Key Laboratory of Survey Science, Yunnan University, Kunming, Yunnan 650500, People's Republic of China}
\email[]{fangyuan@ynu.edu.cn}

\author[]{Chenlei Guo}
\affiliation{National Astronomical Observatories, Chinese Academy of Sciences, Datun Road A20, Beijing 100101, China}
\email[]{guocl@bao.ac.cn}

\author[0009-0002-7851-3706]{Dingfang Hu}
\affiliation{Purple Mountain Observatory, Chinese Academy of Sciences, Nanjing 210008, China}
\affiliation{School of Astronomy and Space Sciences, University of Science and Technology of China, Hefei 230026, China}
\email[]{dfhu@pmo.ac.cn}

\author[0009-0009-3619-6790]{Jie Hu}
\affiliation{Purple Mountain Observatory, Chinese Academy of Sciences, Nanjing 210008, China}
\affiliation{School of Astronomy and Space Sciences, University of Science and Technology of China, Hefei 230026, China}
\email[]{hujie@pmo.ac.cn}

\author[0009-0005-2787-9152]{Qiuju Huang}
\affiliation{Purple Mountain Observatory, Chinese Academy of Sciences, Nanjing 210008, China}
\affiliation{School of Astronomy and Space Sciences, University of Science and Technology of China, Hefei 230026, China}
\email[]{qjhuang@pmo.ac.cn}

\author[0009-0000-2190-6600]{Yanlong Hua}
\affiliation{Purple Mountain Observatory, Chinese Academy of Sciences, Nanjing 210008, China}
\affiliation{School of Astronomy and Space Sciences, University of Science and Technology of China, Hefei 230026, China}
\email[]{ylhua@pmo.ac.cn}

\author[0000-0001-7199-2906]{Yongfeng Huang}
\affiliation{School of Astronomy and Space Science, Nanjing University, Nanjing 210023, People’s Republic of China}
\affiliation{Key Laboratory of Modern Astronomy and Astrophysics (Nanjing University), Ministry of Education, People’s Republic of China}
\email[]{hyf@nju.edu.cn}

\author[0009-0006-7516-2647]{HongGuang Li}
\affiliation{South-Western Institute for Astronomy Research, Yunnan University, Kunming, Yunnan 650500, People's Republic of China}
\affiliation{Yunnan Key Laboratory of Survey Science, Yunnan University, Kunming, Yunnan 650500, People's Republic of China}
\email[]{lihg@stu.ynu.edu.cn}

\author[0000-0003-0394-1298]{Xiangkun Liu}
\affiliation{South-Western Institute for Astronomy Research, Yunnan University, Kunming, Yunnan 650500, People's Republic of China}
\affiliation{Yunnan Key Laboratory of Survey Science, Yunnan University, Kunming, Yunnan 650500, People's Republic of China}
\email[]{liuxk@ynu.edu.cn}

\author[0000-0003-1295-2909]{Xiaowei Liu}
\affiliation{South-Western Institute for Astronomy Research, Yunnan University, Kunming, Yunnan 650500, People's Republic of China}
\affiliation{Yunnan Key Laboratory of Survey Science, Yunnan University, Kunming, Yunnan 650500, People's Republic of China}
\email[]{x.liu@ynu.edu.cn}

\author[]{Yuan Liu}
\affiliation{National Astronomical Observatories, Chinese Academy of Sciences, Datun Road A20, Beijing 100101, China}
\email[]{liuyuan@bao.ac.cn}

\author[]{Jirong Mao}
\affiliation{Yunnan Observatories, Chinese Academy of Sciences, 650011 Kunming, Yunnan Province, People’s Republic of China}
\email[]{jirongmao@ynao.ac.cn}

\author[0009-0009-8313-1842]{Xuan Mao}
\affiliation{National Astronomical Observatories, Chinese Academy of Sciences, Datun Road A20, Beijing 100101, China}
\affiliation{School of Astronomy and Space Sciences, University of Chinese Academy of Sciences, Datun Road A20, Beijing 100049, China}
\affiliation{INAF - Osservatorio Astronomico di Palermo, Piazza del Parlamento 1, 90134 Palermo, Italy}
\email[]{maoxuan@bao.ac.cn}

\author[0000-0003-3635-5375]{Bao Wang}
\affiliation{Purple Mountain Observatory, Chinese Academy of Sciences, Nanjing 210008, China}
\affiliation{School of Astronomy and Space Sciences, University of Science and Technology of China, Hefei 230026, China}
\email[]{baowang@pmo.ac.cn}

\author[0000-0001-6374-8313]{Yuanpei Yang}
\affiliation{South-Western Institute for Astronomy Research, Yunnan University, Kunming, Yunnan 650500, People's Republic of China}
\affiliation{Yunnan Key Laboratory of Survey Science, Yunnan University, Kunming, Yunnan 650500, People's Republic of China}
\email[]{ypyang@ynu.edu.cn}

\author[0000-0003-4121-5684]{Shiyan Zhong}
\affiliation{South-Western Institute for Astronomy Research, Yunnan University, Kunming, Yunnan 650500, People's Republic of China}
\affiliation{Yunnan Key Laboratory of Survey Science, Yunnan University, Kunming, Yunnan 650500, People's Republic of China}
\email[]{zhongsy@ynu.edu.cn}

\author[0009-0008-5929-6658]{Dan Zhu}
\affiliation{South-Western Institute for Astronomy Research, Yunnan University, Kunming, Yunnan 650500, People's Republic of China}
\affiliation{Yunnan Key Laboratory of Survey Science, Yunnan University, Kunming, Yunnan 650500, People's Republic of China}
\email[]{zhud@stu.ynu.edu.cn}

\author[0000-0002-6299-1263]{Xuefeng Wu}
\affiliation{Purple Mountain Observatory, Chinese Academy of Sciences, Nanjing 210008, China} 
\affiliation{School of Astronomy and Space Sciences, University of Science and Technology of China, Hefei 230026, China} 
\email[show]{xfwu@pmo.ac.cn}

\author[]{Weimin Yuan}
\affiliation{National Astronomical Observatories, Chinese Academy of Sciences, Datun Road A20, Beijing 100101, China}
\email[]{wmy@nao.cas.cn}

\correspondingauthor{Di Xiao, Jinjun Geng, Xuefeng Wu} 

\begin{abstract}
GRB 251222A is a long-duration gamma-ray burst at a spectroscopic redshift of $z=3.171$. It was detected by the \EP\ as the soft X-ray transient EP251222b before and during the gamma-ray emission. We performed a multi-wavelength (\EP/WXT, \fermi/GBM, \swift/XRT and optical/NIR) investigation of this event. Its prompt soft X-ray and gamma-ray features, together with two afterglow rebrightenings, allow us to examine a possible connection between the prompt ejecta and the afterglow. We consider a multiple-shell scenario in which a faster, narrower shell catches up with a slower, wider shell. We assume that the selected pulse groups trace different ejecta and that the late X-ray flare marks their collision. Under these assumptions, we represent the post-collision outflow by an effective merged core and an outer annular component. A joint fit to the multi-band afterglow gives a Lorentz factor of $\Gamma_c\approx293$ for the merged core and $\Gamma_{1,w}\approx228$ for the outer component. The sample-wise collision mapping gives an energy release of $E_{\rm diss}=(1.6^{+2.1}_{-0.9})\times10^{52}$ erg and a dynamical dissipation efficiency of $0.37^{+0.07}_{-0.14}\%$. EP251222b can therefore be used to test a prompt--afterglow connection through shell collisions, although it does not uniquely establish this interpretation.
\end{abstract}


\section{Introduction}\label{sec:introduction}

Gamma-ray bursts (GRBs) are among the most luminous transient phenomena in the Universe, serving as vital subjects for research in time-domain astronomy and high-energy astrophysics. GRBs release tremendous amounts of energy in the form of gamma-rays associated with the collapse of massive stars or the merger of compact objects \citep{EichlerLivio1989,Kouveliotou1993,KluzniakRuderman1998,WoosleyBloom2006,Berger2014}, where the prompt emission is believed to be produced by internal shocks or magnetic reconnection in the relativistic jets launched by the central engine \citep{BlandfordRoger1977,SariPiran1997,Kobayashi1997,Piran2004,ZhangYan2011}. After the prompt emission, the interaction between the jet and the surrounding medium produces a long-lasting afterglow emission across the electromagnetic spectrum \citep{BlandfordMckee1976,MeszarosRees1997,SariPiranNarayan1998,SariPiran1999,Huang2000,Piran2004,WangZhang2015}, which can be used to probe the physical properties of the jet and the circumburst environment.

The simplest forward-shock model predicts smooth power-law evolution, but many afterglows contain complex features like flares, plateaus, and rebrightenings. Proposed explanations include refreshed shocks caused by late shells \citep{ReesMeszaros1998,KumarPiran2000,SariMeszaros2000,GranotNakar2003}, late energy injection from a magnetar central engine \citep{Dai1998,ZhangMeszaros2001}, angularly structured or two-component jets \citep{Rossi2002,ZhangMeszaros2002,Granot2002}, density variations in the external medium \citep{NakarGranot2007}, and late internal dissipation \citep{FanWei2005}. A useful discriminator is whether the prompt pulse sequence, late high-energy activity, and afterglow peaks can be described by one dynamical history rather than by unrelated components.

The \ep\ mission, launched on 2024 January 9, is particularly well suited to this problem because its Wide-field X-ray Telescope (WXT) monitors the soft X-ray sky while the Follow-up X-ray Telescope (FXT) provides deeper pointed observations \citep{Yuan2022}. EP251222b was detected by WXT in temporal and spatial association with GRB~251222A \citep{GCN43201,GCN43188,GCN43186}. The event was subsequently localized across X-ray and optical wavelengths and assigned a spectroscopic redshift of $z=3.171$ \citep{GCN43204}. Its combination of multiple prompt episodes, very early optical detections, an X-ray flare, and two optical afterglow bumps makes it a useful test case for time-dependent jet structure.

Recent work has begun to connect complex prompt activities and the late-time afterglow structure within a common evolution of multiple ejecta. High-energy variability and structure in the later afterglow may arise from the same dynamical sequence if the central engine releases relativistic material more than once. A newly launched outflow can produce high-energy emission before or during its encounter with earlier ejecta, whereas the redistribution of energy and momentum in that encounter subsequently modifies the external-shock emission. GRB~240529A showed two prompt episodes separated by a quiescent interval of about 400~s, followed by an apparent X-ray plateau and optical rebrightening \citep{Svinkin2024}. By combining the prompt analysis with numerical modeling of the optical afterglow, \citet{Sun2024} interpreted these features as two distinct shocks launched during separate episodes of central-engine activity. GRB~060729 provided a more explicit example of this connection: its prompt pulses and two X-ray flares were followed by two broad bumps in the X-ray and optical afterglows. Using the timing of the prompt pulses and flares together with multiwavelength afterglow modeling, \citet{Geng2025} identified three ejecta shells whose successive collisions produced the X-ray flares, while the resulting fast narrow component and slower hollow component generated the two later afterglow bumps.

Related interpretations have also been applied to the multi-epoch rebrightenings of GRB~250129A \citep{Akl2026} where numerical calculations reproduced the observations with successive refreshed shocks caused by delayed shell collisions. The work on EP250302a showed a sharp X-ray flare followed by a smoother optical and X-ray rebrightening and linked both features to a second shell launched by the late central-engine reactivation and its subsequent collision with the leading ejecta \citep{Fu2026}.

EP251222b/GRB~251222A presents a closely related set of observables. Therefore, in this work, we use the prompt timing to construct a shell-collision mapping and test whether the resulting effective core--wing outflow can reproduce the multi-band afterglow. The purpose of this calculation is to quantify a physically allowed connection between the prompt emission and the afterglow.

We present the observations and data analysis in Section~\ref{sec:obs}, introduce a shell-collision scenario in Section~\ref{sec:model}, and describe the fitting strategy and results in Section~\ref{sec:result}. Section~\ref{sec:discussion} discusses alternative interpretations, model limitations, and testable consequences, and Section~\ref{sec:conclusion} summarizes the main conclusions.

\section{Observations and Data Analysis}\label{sec:obs}

\subsection{Prompt emission}\label{sec:prompt}

EP251222b triggered WXT onboard \EP\ at 17:05:34 (UTC dates and times are used throughout the paper) on 2025 December 22. The WXT position was $\mathrm{R.A.}=77.207^\circ$ and $\mathrm{Dec.}=-7.208^\circ$ (J2000), with a 3 arcmin radius uncertainty \citep{GCN43201}. Our refined WXT analysis indicates that weak but detectable emission began at 2025-12-22 17:04:52 UTC, approximately 42 s before the onboard trigger time. We denote the emission onset by $T_{0,\rm EP}$ and adopt $T_0\equiv T_{0,\rm EP}$ for the prompt light curves, prompt tables, and follow-up epochs. The WXT onboard trigger is therefore at $T_0+42$ s. In the shell-timing calculation in Section~\ref{sec:model}, we instead measure time from the Fermi/GBM trigger, $T_{0,\rm GBM}$. The two broadband model-comparison figures use the ECLAIRS plotting zero, $T_{0,\rm ECL}$; the corresponding time conversions are given in their captions. The event remained detectable in the 0.5--4 keV band for approximately 350 s. We modeled the time-averaged WXT spectrum over the 350 s emission interval with an absorbed power law,
\begin{equation}
N(E)=\exp\left(-N_{\rm H}\sigma(E)\right)K E^{-\alpha_{\rm APL}},
\label{eq:apl}
\end{equation}
where $N_{\rm H}$ is the equivalent hydrogen column density and $\sigma(E)$ is the photoelectric absorption cross section. With the Galactic absorption fixed at $9.81\times10^{20}\ \mathrm{cm^{-2}}$, the photon index is $1.05\pm0.15$, and the average unabsorbed 0.5--4 keV flux is $2.8^{+0.3}_{-0.3}\times10^{-9}\ \mathrm{erg\,cm^{-2}\,s^{-1}}$. This measurement is distinct from the WXT quantities in Table~\ref{tab:properties}, which were derived over the WXT $T_{90}$ interval.

\fermi/GBM was triggered at 2025-12-22 17:05:51.27 UTC, corresponding to $T_{0,\rm GBM}=T_{0,\rm EP}+59.27$ s and 17.27 s after the WXT onboard trigger \citep{Meegan2009,GCN43186}. The GBM duration ($T_{90}$) is about 83 s in the 10--1000 keV band \citep{GCN43222}. A CPL fit to the time-averaged spectrum from $T_{0,\rm GBM}-2.2$ to $T_{0,\rm GBM}+91.7$ s gives a photon index of $1.26^{+0.01}_{-0.01}$ and a peak energy of $118^{+1}_{-1}\,\rm{keV}$. A Band model also provides a good fit, with a low-energy photon index of $-0.84^{+0.01}_{-0.01}$, a high-energy photon index of $-2.00^{+0.02}_{-0.02}$, and a peak energy of $64.9^{+0.8}_{-0.8}\,\rm{keV}$. The derived 10--1000 keV fluence is $1.09^{+0.02}_{-0.02} \times 10^{-5} \, \mathrm{erg\, cm^{-2}}$ \citep{GCN43222}. The GBM Circular gives the signed power-law exponent as $-1.26$. In our CPL and PL equations, we instead define a positive photon index through $N(E)\propto E^{-\alpha}$. Table~\ref{tab:properties} gives our fit over the instrument-specific $T_{90}$ interval, which differs from the wider interval used in the Circular. The two sets of spectral parameters should therefore not be compared directly. The burst was also detected by \svom, \textit{Konus-Wind}, and \textit{GECAM-B} \citep{GCN43188,GCN43214,GCN43218}.

\begin{figure}[t]
\centering
\includegraphics[width=1\columnwidth]{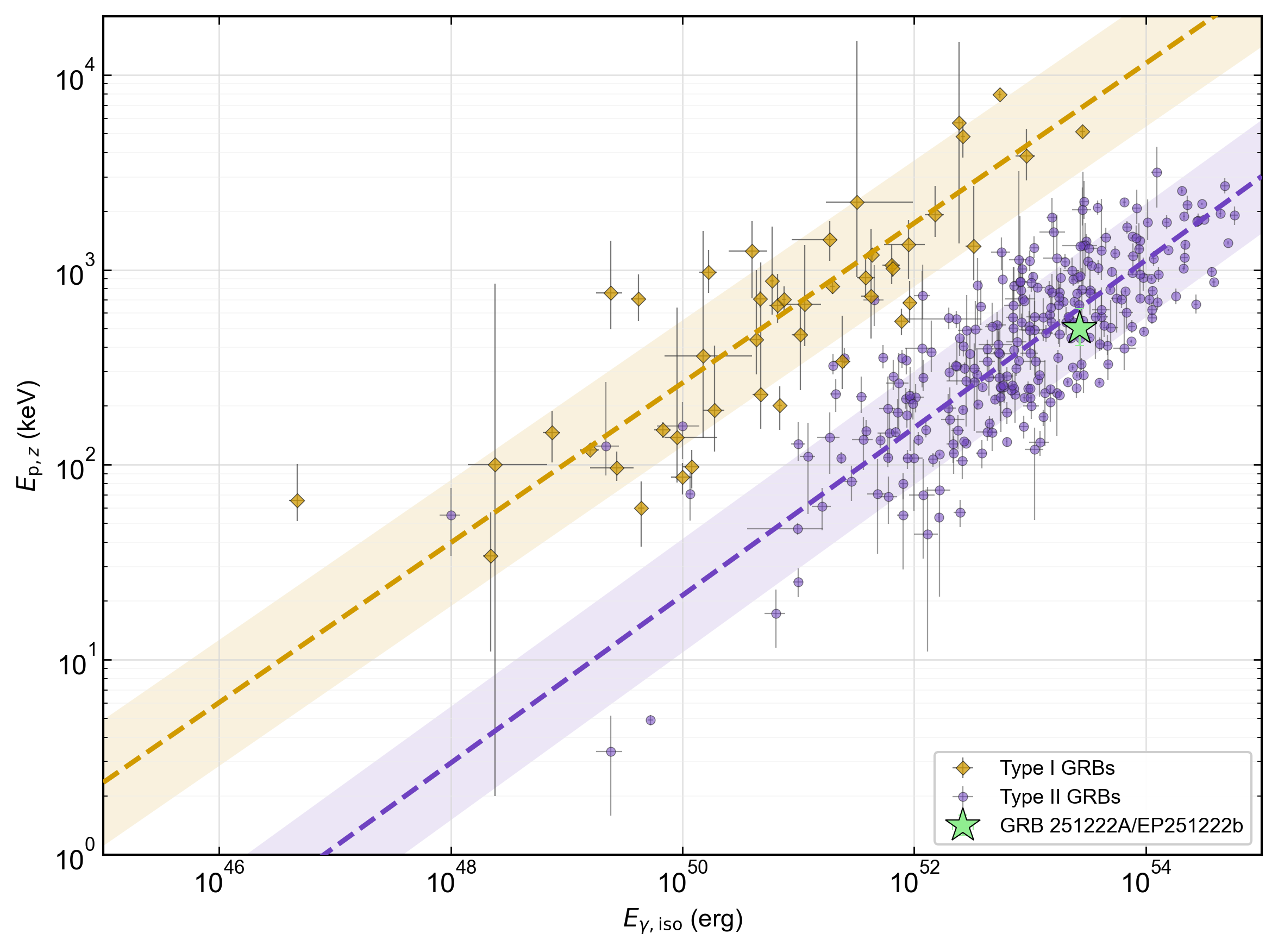}
\caption{Location of EP251222b/GRB 251222A (green star) in the
$E_{\rm p,z}$--$E_{\gamma,\rm iso}$ plane. The background population of typical Type II GRBs (massive star GRBs) is shown as purple points, while Type I GRBs (compact star GRBs) are shown as yellow points \citep{Minaev2020}. The shaded regions indicate the $1\sigma$ confidence intervals. The green-star error bars show the conditional 90\% confidence intervals from the CPL fit to data from three Fermi/GBM detectors, with $E_{\gamma,\rm iso}$ integrated over the $T_{90}$ interval and the rest-frame 1--10000 keV band. 
\label{fig:amati}}
\end{figure}

We calculated $T_{90}$ from the background-subtracted 0.5--4 keV WXT light curve and the 8--900 keV GBM NaI light curve as the interval containing 5--95\% of the cumulative counts. The isotropic-equivalent radiated energy was computed from the bolometric fluence $S_{\rm bolo}$ as
\begin{equation}
E_{\rm iso}=\frac{4\pi D_L^2 S_{\rm bolo}}{1+z},
\label{eq:eiso}
\end{equation}
where $D_L$ is the luminosity distance and the $K$-correction factor is applied following \citet{Bloom2001}. We adopt the Planck18 cosmology implemented in \texttt{Astropy} \citep{Astropy2013,Astropy2018,Astropy2022}. With a spectroscopic redshift of $z=3.171$, we derive the isotropic energy in the gamma-ray band and soft X-ray band. The rest-frame peak energy is $E_{\rm{p,z}}=E_{\rm{p}}(1+z)$. As shown in Figure~\ref{fig:amati}, the location of GRB 251222A is consistent with the long-GRB branch of the $E_{\rm p,z}$--$E_{\gamma,\rm iso}$ relation (the ``Amati relation''; \citet{Amati2002,Minaev2020}; for a discussion of potential selection effects in this relation, see \citet{Butler2007}). Together with its long duration, this burst is consistent with a Type II (massive-star) origin.
GRB~251222A also lies at the position of typical bright GRBs in the Amati plane, which suggests that this burst is likely viewed on-axis (e.g., \citet{GaoGeng2025}).
A summary of the basic observed properties of GRB 251222A is listed in Table \ref{tab:properties}.

\begin{deluxetable}{ccc}
\tablewidth{\columnwidth}
\tablecaption{Observed Gamma-ray and X-ray Properties of GRB
251222A/EP251222b.
\label{tab:properties}}
\tablehead{
\colhead{Property} & \colhead{\fermi/GBM} & \colhead{\ep/WXT}
}
\startdata
$T_{90}$ (s) & $82.4^{+0.8}_{-1.3}$ & $225.1^{+8.2}_{-7.3}$\\
Photon index & $1.4^{+0.2}_{-0.2}$ & $1.1^{+0.2}_{-0.2}$\\
Peak energy (keV) & $119.0^{+72.3}_{-31.7}$ & \nodata\\
Rest-frame peak energy (keV) & $496.5^{+301.6}_{-132.3}$ & \nodata\\
Band-limited Fluence\tablenotemark{a} ($\mathrm{erg\,cm^{-2}}$) &
$1.1^{+0.2}_{-0.2}\times10^{-5}$ & $1.2^{+0.1}_{-0.1}\times10^{-6}$\\
Band-limited $E_{\rm iso}$ (erg) & $2.4^{+0.4}_{-0.3}\times10^{53}$ & $2.6^{+0.3}_{-0.3}\times10^{52}$\\
\enddata
\tablecomments{$^{\rm{a}}$The X-ray and gamma-ray fluences are calculated in the 0.5--4 keV and 8--900 keV bands, respectively. All quantities are derived from the time-averaged spectra over the corresponding instrument-specific $T_{90}$ intervals. The $T_{90}$ uncertainties denote the central 68\% intervals from Monte Carlo resampling, whereas spectral-parameter uncertainties denote 90\% confidence intervals.}
\end{deluxetable}

Both the soft X-ray and gamma-ray light curves contain multiple emission episodes. While the X-ray emission is longer, there are two additional episodes before and after the gamma-ray emission. Based on this multi-peak profile, we divided the prompt interval into six time slices and performed joint fits to \ep/WXT, two NaI detectors, and one BGO detector of \fermi/GBM. The joint spectra are well modeled by a cutoff power-law (CPL) or a power-law (PL) function with absorption included for the WXT component,
\begin{align}
N(E)&=KE^{-\alpha_{\rm{CPL}}}\exp\left( -\frac{E}{E_{\mathrm{cut}}} \right), \label{eq:cpl}\\
N(E)&=KE^{-\alpha_{\rm PL}}, \label{eq:pl}
\end{align}
where $\alpha$ is the photon index and for the cutoff power law, $E_{\rm{cut}}$ is the cutoff energy. The peak energy of the $\nu F_{\nu}$ spectrum is given by $E_{\rm{p}}=(2-\alpha_{\rm{CPL}})E_{\rm{cut}}$. 
We also fitted the Band function \citep{Band1993},
\begin{equation}
\begin{aligned}
N_{\rm B}(E)&=K_{\rm B}
\begin{cases}
(E/E_{\rm piv})^{\alpha_{\rm B}}e^{-E/E_0}, & E\leq E_{\rm br},\\[1mm]
C_{\rm B}(E/E_{\rm piv})^{\beta_{\rm B}}, & E>E_{\rm br},
\end{cases}\\[2mm]
C_{\rm B}&=(E_{\rm br}/E_{\rm piv})^{\alpha_{\rm B}-\beta_{\rm B}}e^{\beta_{\rm B}-\alpha_{\rm B}},
\end{aligned}
\label{eq:band}
\end{equation}
where $E_{\rm piv}=100$ keV and $E_{\rm br}=(\alpha_{\rm B}-\beta_{\rm B})E_0$. Here $\alpha_{\rm B}$ and $\beta_{\rm B}$ are signed exponents. The peak energy is then given by $E_{\rm{p}}=(2+\alpha_{\rm B})E_0$. 

The uncertainties in both peak energies is determined through: 
\begin{equation}
\Delta S(E_{\rm p})=
\min_{\boldsymbol{\theta}:\,f(\boldsymbol{\theta})=E_{\rm p}}
S(\boldsymbol{\theta})-S_{\rm min}.
\end{equation}
At each trial $E_{\rm p}$, the remaining free parameters are re-optimized within their allowed ranges. The 90\% confidence limits correspond to an increase of $\Delta S=2.706$ above the minimum fit statistic. The rest-frame energy and its uncertainties are multiplied by $1+z$, with $z$ fixed. Formula-derived energies in Table~\ref{tab:timeresolved_spectra} are retained without imposing additional physical-peak conditions.
For the spectral envelopes in Figures~\ref{fig:jspec} and \ref{fig:bbsed}, we estimate the uncertainties in the unabsorbed spectra using the full best-fit covariance matrix, including correlations between parameters, with errors propagated in logarithmic flux. The shaded bands show approximate 90\% confidence intervals at each energy for the adopted model and fixed absorption. This local Gaussian approximation may be less reliable for poorly constrained spectra or extrapolation far beyond the fitted energy range.

\begin{figure}[t]
\centering
\includegraphics[width=\columnwidth]{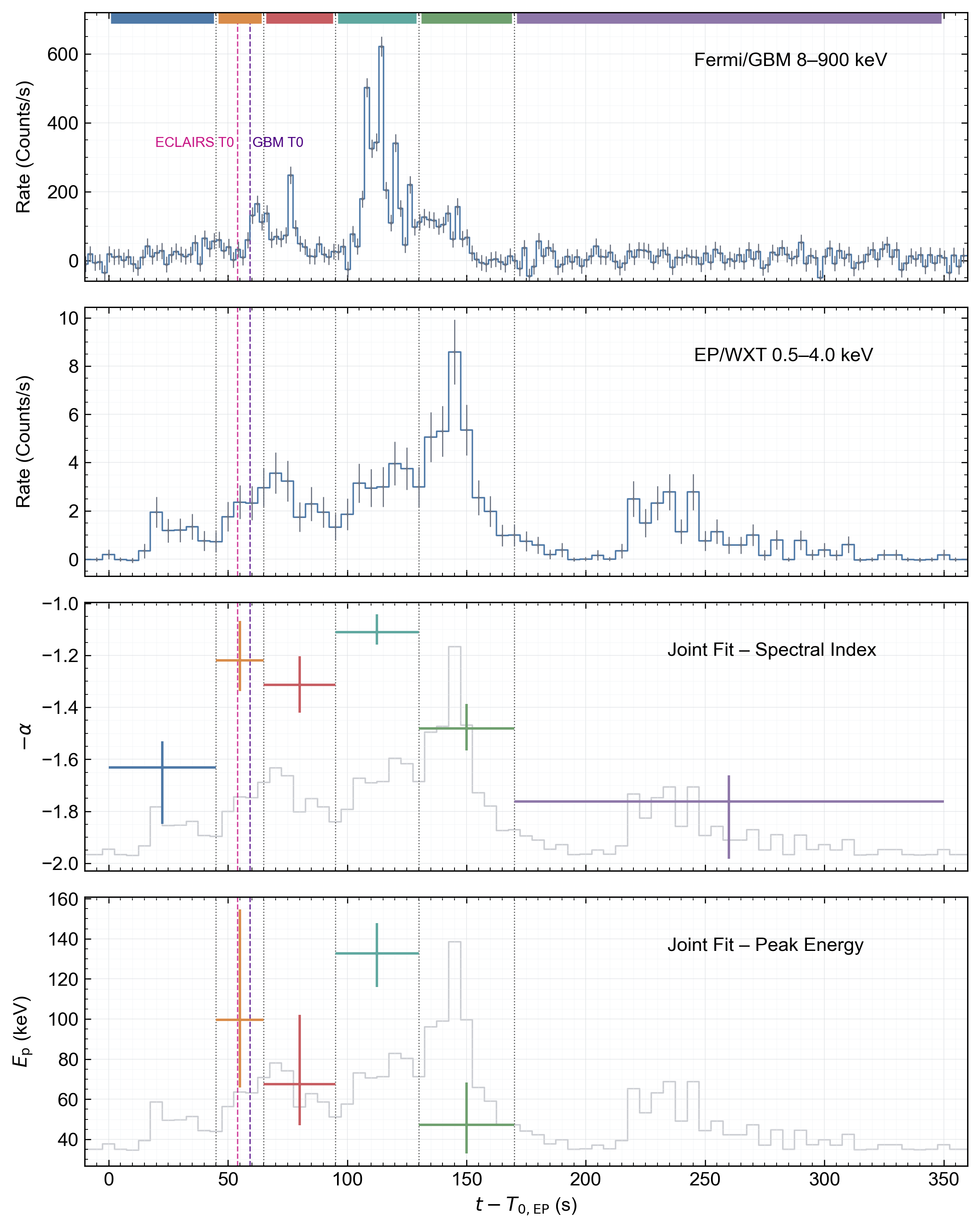}
\caption{The light curves and the time-resolved joint spectral fit of \ep/WXT and 3 \fermi/GBM detectors (NaI-10, NaI-11, BGO-01). The EP/WXT data are displayed with a bin size of 5 seconds and the \fermi/GBM data with a bin size of 2 seconds. If there is no peak energy data point in the bottom panel, it means that the cutoff energy cannot be constrained, and the PL model is therefore adopted. The vertical lines separate the time intervals we choose. Time is measured from $T_{0,\rm EP}$ = 2025 December 22 17:04:52 UTC.
\label{fig:jlc}}
\end{figure}

\begin{figure}[t]
\centering
\includegraphics[width=\columnwidth]{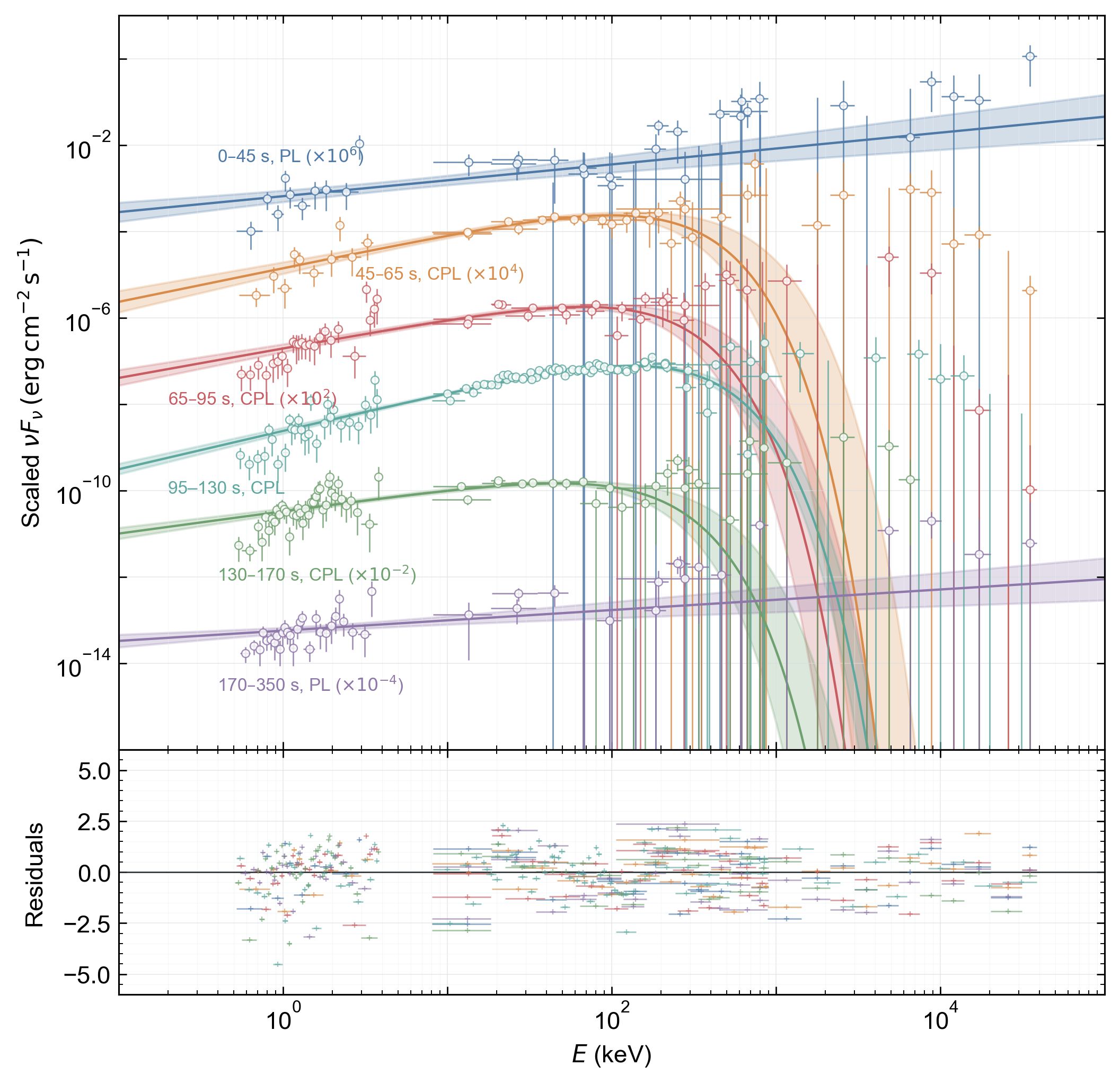}
\caption{The time-resolved spectra from the joint fits of \ep/WXT and 3 \fermi/GBM detectors. The earliest and the latest are fitted with a PL model, while the others are fitted with a CPL model. The color is the same as the time intervals in Figure \ref{fig:jlc}. The solid lines are the unabsorbed models and the shaded regions denote pointwise 90\% confidence envelopes, estimated from the covariance matrix under a local Gaussian approximation. The residuals of all the data points are shown in the bottom panel.
\label{fig:jspec}}
\end{figure}

For the GBM dataset, we selected the  NaI-10, NaI-11, and BGO-01 detectors. We used 8--900 keV for the NaI detectors and 100--40000 keV for the BGO detector, and reduced the data with the Gamma-ray Data Tools \citep{GDT-Core,GDT-Fermi} to extract energy-resolved light curves and spectra. For the WXT data, we performed data reduction following standard procedures within the \texttt{HEAsoft} package \citep{HEAsoft}. 
All spectral modeling was conducted using the XSPEC software package (v12.15; \citet{Arnaud1996,DormanArnaud2001,DormanArnaudGordon2003}). For the GBM data, the background spectra were estimated from the $[-130,-100]$ and $[300,470]$ s intervals relative to $T_{0,\rm GBM}$ and modeled with a first-order polynomial over time. The GBM spectra were retained at their native channel grouping, while the WXT spectra were grouped to contain at least one count per bin. We used PGSTAT for the NaI and BGO spectra and CSTAT for the WXT spectrum in XSPEC. We initially allowed multiplicative constants to vary between instruments to account for possible cross-calibration offsets. However, these constants were not constrained by the time-resolved spectra; the fits either became unstable or drove the constants to the imposed boundaries, even when restricted ranges were adopted. We therefore fixed the relative cross-normalization constants to unity. We use six joint-fit intervals: $T_0+[0,45]$, $[45,65]$, $[65,95]$, $[95,130]$, $[130,170]$, and $[170,350]$ s, where $T_0=T_{0,\rm EP}$. In Figures~\ref{fig:jlc} and \ref{fig:jspec}, the negative spectral exponent corresponds to the positive photon index $\alpha$ defined in Equations~(\ref{eq:cpl}) and (\ref{eq:pl}).

Figures~\ref{fig:jlc} and \ref{fig:jspec} show the prompt light curves and the six time-resolved spectra, respectively, while Table~\ref{tab:timeresolved_spectra} summarizes the corresponding PL, CPL and Band fits.
For the earliest ($T_0+[0,45]$ s) and latest ($T_0+[170,350]$ s) intervals, the cutoff energy cannot be constrained whatever the adopted upper bound is and the characteristic energy for Band model is also poorly constrained; we therefore adopt the simpler PL model. The corresponding values of $\Delta{\rm BIC}={\rm BIC}_{\rm PL}-{\rm BIC}_{\rm CPL}$ are $-5.51$ and $-6.78$, respectively, which also favor PL in these two intervals. To describe the spectral evolution with a consistent parameterization, the CPL model is adopted for the four central intervals whenever a finite CPL solution can be obtained.
We report all three kinds of fits in Table~\ref{tab:timeresolved_spectra}, because the BIC is used as a diagnostic rather than as the sole model-selection criterion. In particular, although the BIC favors Band in the $T_0+[95,130]$ s interval, we retain its finite CPL solution to maintain a consistent spectral parameterization across the four central intervals and use the Band fits to assess the sensitivity of the inferred spectral properties to model choice.

Compared with the preceding interval under the CPL parameterization, the brightest gamma-ray interval at $T_0+[95,130]$ s has a harder best-fitting index and a higher peak energy: the best-fitting photon index decreases from $\alpha=1.31^{+0.11}_{-0.11}$ to $\alpha=1.11^{+0.05}_{-0.07}$. Treating the two measurements as independent and approximately Gaussian, and converting their 90\% uncertainties to $1\sigma$ uncertainties, gives an estimated hardening significance of approximately $2.7\sigma$. We therefore regard the hardening as suggestive.
Moreover, the first two gamma-ray pulses overlap, and we have not performed a quantitative pulse-width analysis. Thus, associating the earlier pulse group with one shell and the later hard pulse with another remains a model assumption. A soft X-ray-dominated flare subsequently appears near $T_0+220$ s, after the main gamma-ray activity.

\begin{deluxetable*}{ccccccccc}
\tablewidth{\textwidth}
\tablecaption{Time-resolved joint spectral fits of EP/WXT and Fermi/GBM data.\label{tab:timeresolved_spectra}}
\tablehead{
\colhead{Interval ($T-T_0$)} & \colhead{Model} & \colhead{$\alpha/\alpha_{\rm B}$} &
\colhead{$\beta_{\rm B}$} & \colhead{$E_{\rm cut}/E_0$} & \colhead{$E_{\rm p}$} &
\colhead{Statistic/dof} & \colhead{AIC} & \colhead{BIC}
}
\startdata
$0$--$45$ s & PL$^{\rm a}$ & $1.63^{+0.22}_{-0.10}$ & \nodata & \nodata & \nodata & $429.82/390$ & $433.82$ & $441.76$ \\
 & CPL & $1.57^{+0.22}_{-0.51}$ & \nodata & $341^{+\mathrm{NC}}_{-341}\,^{\dagger}$ & $145^{+126}_{-136}$ & $429.37/389$ & $435.37$ & $447.28$ \\
 & Band & $-1.08^{+0.23}_{-0.58}\,^{\dagger}$ & $-1.89^{+0.30}_{-0.82}\,^{\dagger}$ & $10^{+\mathrm{NC}}_{-\mathrm{NC}}\,^{\dagger}$ & $9^{+4302}_{-6}$ & $428.18/388$ & $436.18$ & $452.07$ \\
$45$--$65$ s & PL & $1.53^{+0.05}_{-0.06}$ & \nodata & \nodata & \nodata & $382.36/390$ & $386.36$ & $394.31$ \\
 & CPL$^{\rm a}$ & $1.22^{+0.12}_{-0.15}$ & \nodata & $128^{+96}_{-54}$ & $100^{+55}_{-34}$ & $357.00/389$ & $363.00$ & $374.92$ \\
 & Band & $-1.14^{+0.17}_{-0.14}\,^{\dagger}$ & $-2.10^{+0.34}_{-\mathrm{NC}}\,^{\dagger}$ & $81^{+70}_{-64}$ & $69^{+44}_{-49}$ & $354.54/388$ & $362.54$ & $378.43$ \\
$65$--$95$ s & PL & $1.61^{+0.04}_{-0.05}$ & \nodata & \nodata & \nodata & $446.19/431$ & $450.19$ & $458.33$ \\
 & CPL$^{\rm a}$ & $1.31^{+0.11}_{-0.11}$ & \nodata & $98^{+71}_{-37}$ & $67^{+35}_{-20}$ & $415.82/430$ & $421.82$ & $434.03$ \\
 & Band & $-1.20^{+0.19}_{-0.16}$ & $-2.07^{+0.19}_{-1.06}$ & $45^{+62}_{-34}\,^{\dagger}$ & $36^{+35}_{-20}$ & $412.73/429$ & $420.73$ & $437.01$ \\
$95$--$130$ s & PL & $1.52^{+0.01}_{-0.02}\,^{\dagger}$ & \nodata & \nodata & \nodata & $737.45/443$ & $741.45$ & $749.64$ \\
 & CPL$^{\rm a}$ & $1.11^{+0.05}_{-0.07}$ & \nodata & $149^{+24}_{-26}$ & $133^{+15}_{-17}$ & $459.41/442$ & $465.41$ & $477.71$ \\
 & Band & $-1.01^{+0.07}_{-0.08}$ & $-2.03^{+0.10}_{-0.20}$ & $92^{+26}_{-17}$ & $91^{+18}_{-14}$ & $446.08/441$ & $454.08$ & $470.47$ \\
$130$--$170$ s & PL & $1.73^{+0.04}_{-0.04}$ & \nodata & \nodata & \nodata & $482.98/476$ & $486.98$ & $495.31$ \\
 & CPL$^{\rm a}$ & $1.48^{+0.08}_{-0.09}$ & \nodata & $91^{+59}_{-35}$ & $47^{+21}_{-14}$ & $452.75/475$ & $458.75$ & $471.26$ \\
 & Band & $-1.46^{+0.11}_{-0.09}\,^{\dagger}$ & $-2.45^{+0.36}_{-\mathrm{NC}}\,^{\dagger}$ & $77^{+52}_{-24}\,^{\dagger}$ & $41^{+19}_{-19}$ & $451.20/474$ & $459.20$ & $475.88$ \\
$170$--$350$ s & PL$^{\rm a}$ & $1.76^{+0.22}_{-0.10}$ & \nodata & \nodata & \nodata & $510.75/443$ & $514.75$ & $522.95$ \\
 & CPL & $1.74^{+0.24}_{-0.22}$ & \nodata & $\ddagger$ & $128^{+60}_{-120}\,^{\dagger}$ & $511.43/442$ & $517.43$ & $529.73$ \\
 & Band & $-1.76^{+0.38}_{-0.08}\,^{\dagger}$ & $-1.75^{+0.12}_{-0.41}\,^{\dagger}$ & $286^{+\mathrm{NC}}_{-\mathrm{NC}}\,^{\dagger}$ & $69^{+3376}_{-68}$ & $510.77/441$ & $518.77$ & $535.16$ \\
\enddata
\tablecomments{The energies are all in keV, and parameter uncertainties are quoted at the 90\% confidence level.
$\alpha$ is the positive photon index of PL/CPL, while $\alpha_{\rm B}$ and $\beta_{\rm B}$ are the native Band exponents.
The energy column gives $E_{\rm cut}$ for CPL and $E_0$ for Band.
$E_{\rm p}=(2-\alpha)E_{\rm cut}$ for CPL and $E_{\rm p}=(2+\alpha_{\rm B})E_0$ for Band; its uncertainties come from profiling the corresponding formula with $\Delta S=2.706$.
$^{\rm a}$Adopted model; $\dagger$XSPEC error-search warning; NC denotes the parameter uncertainty could not be constrained; $\ddagger$The best-fitting parameter reaches its boundary.
For $\Delta{\rm BIC}={\rm BIC}_{\rm PL}-{\rm BIC}_{\rm CPL}$, the six intervals give $-5.51$, $19.39$, $24.30$, $271.93$, $24.06$, $-6.78$, respectively; for $\Delta{\rm BIC}={\rm BIC}_{\rm BAND}-{\rm BIC}_{\rm CPL}$, the six intervals give $4.79$, $3.51$, $2.98$, $-7.24$, $4.63$, $5.44$, respectively.}
\end{deluxetable*}

\subsection{Broadband afterglow observations}\label{sec:afterglowobs}

\subsubsection{X-ray follow-up}

\ep/FXT did not perform an auto follow-up since it had been observing another target at the time of the trigger. It began a pointed observation approximately 12.1 hr after $T_0$ and detected an uncatalogued source at $\mathrm{R.A.}=77.1954^\circ$ and $\mathrm{Decl.}=-7.2139^\circ$ (J2000), with an uncertainty of about 10 arcsec \citep{GCN43215}. An absorbed power law fitted to the time-averaged 0.5--10 keV spectrum gives a photon index of $1.86^{+0.26}_{-0.25}$ and an unabsorbed flux of $1.8^{+0.4}_{-0.3}\times10^{-12}\ \mathrm{erg\,cm^{-2}\,s^{-1}}$ with a fixed Galactic equivalent hydrogen column density of $9.81\times 10^{20}\rm{cm^{-2}}$.

\swift\ performed follow-up observations with the X-Ray Telescope (XRT; \citet{Burrows2005swift}). It obtained 2.0 ks of Photon Counting mode data between approximately $T_0+10.7$ and $T_0+17.4$ ks and localized the X-ray afterglow to a 2.5 arcsec radius \citep{GCN43213}. We used the lightcurve and spectrum products provided by the UK Swift Science Data Centre (UKSSDC; \citet{Evans2007,Evans2009}). The spectrum can be fitted with an absorbed power-law with a photon spectral index of $1.92_{-0.20}^{+0.31}$ and a best-fitting absorption column of $1.05_{-0.12}^{+0.91}\times 10^{21}\rm{cm^{-2}}$.

\subsubsection{Optical and near-infrared follow-up}\label{sec:agopt}

We monitored the optical evolution with the \mephisto\ (Multi-channel Photometric Survey Telescope) operated by the South-Western Institute for Astronomy Research (SWIFAR), Yunnan University \citep{Yuan2020}.
This facility has unique capability of simultaneous observations in three bands ($ugi$ or $vrz$). The $u,v,g,r,i,z$ \mephisto\ bands have wavelength coverage of $320-365$, $365-405$, $480-580$, $580-680$, $775-900$, and $900-1050$ nm, respectively \citep[for details on the Mephisto filter system and telescope, see][]{ChenX-2024C, Yuan-PeiY, Guowang2025D, Zou-2026Z, Zhao2026Z}. The multi-band \mephisto\ observations of EP251222b/GRB 251222A field were acquired in two sets, one for $u,g,i$ began at 17:11:08 2025-12-22 ($T_{0}$+6 min) and another for u,g,i,v,r,z began at 17:48:08 2025-12-22 ($T_{0}$+43 min).
All raw data were delivered to the Mephisto Data Center for real-time processing. The images were reduced using a dedicated Mephisto pipeline, including bias, dark subtraction and flat-field correction (Fang et al. in preparation). Astrometric calibration was performed with \href{http://astrometry.net/}{astrometry.net} \citep{Lang2010} and \texttt{SCAMP} \citep{Bertin2006}, using the Gaia DR3 catalog \citep{Gaia2023,Montegriffo2023} as the astrometric reference. Source extraction and aperture photometry were subsequently carried out with \texttt{SExtractor} \citep{BertinArnouts1996}. Photometric calibration was based on synthetic photometry derived from recalibrated Gaia BP/RP (XP) spectra \citep{Gaia2023,Montegriffo2023,Xiao2023,Huang2024}. For each Mephisto band, synthetic AB magnitudes were calculated by convolving the recalibrated XP spectra with the corresponding Mephisto system throughput curve. These throughput curves were constructed from the measured efficiencies of the individual optical components along the light path, from the primary mirror to the detectors. The resulting photometric calibration precisions were approximately 0.02 mag in the \textit{u} and \textit{v} bands, 0.01 mag in the \textit{g} and \textit{r} bands, and below 0.01 mag in the \textit{i} and \textit{z} bands (Zhang et al. in preparation).

\begin{figure*}[hbt!]
\centering
\includegraphics[width=0.9\textwidth]{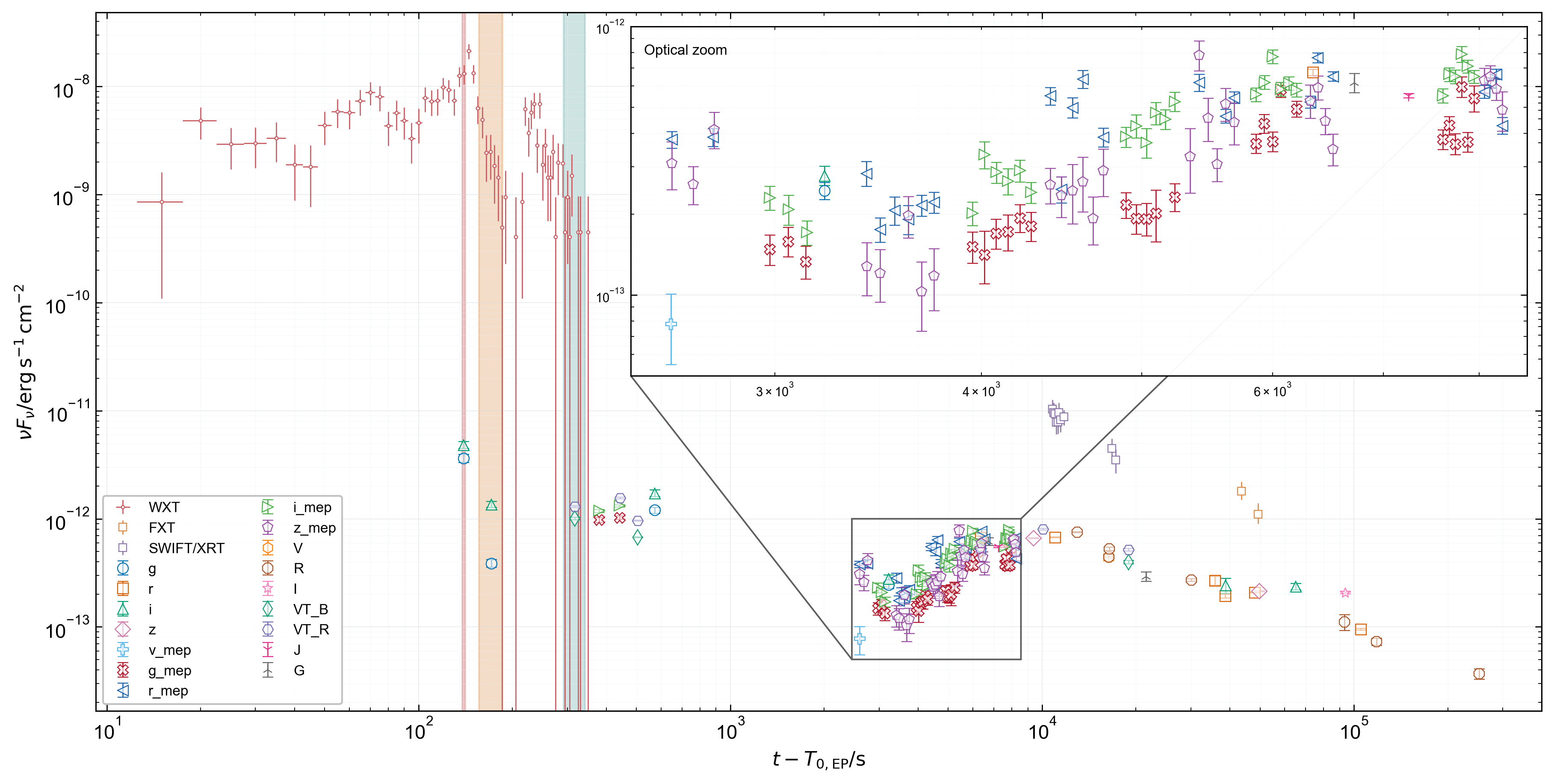}
\caption{X-ray and optical/NIR light curves. Shaded vertical intervals mark the three epochs used for the prompt broadband SED analysis. Labels with the suffix “mep” represent the bands of \mephisto. The optical evolution shows an early bump near several hundred seconds and a broader rebrightening near $10^4$ s.\label{fig:xonlc}}
\end{figure*}

We also compiled optical and near-infrared measurements reported in GCN Circulars\footnote{\url{https://gcn.nasa.gov/circulars}}. The complete optical/NIR photometry, combining the GCN and \mephisto\ measurements, is listed in Appendix~\ref{app:ondata} (Table~\ref{tab:ondata}). Times are mid-exposure times in seconds relative to the adopted \ep\ $T_0$ at 2025 December 22 17:04:52 UTC. Figure~\ref{fig:xonlc} shows the broadband light curve not corrected for any extinction.

To estimate the host extinction while allowing for intrinsic spectral evolution between the prompt and afterglow phases, we fitted the two optical-to-X-ray SEDs independently: one during the prompt-emission phase at $T_0+140$ s, and one during the forward-shock afterglow phase at around $T_0+14$ ks. The optical-to-X-ray SEDs were constructed following the procedures described by \citet{Hu2021} and \citet{CaballeroGarcia2023}, together with the analysis threads provided by the \swift\ team\footnote{\url{https://swift.gsfc.nasa.gov/analysis/threads/uvot_thread_spectra.html}}. For each epoch, the contemporaneous X-ray spectrum and optical/NIR photometric points were fitted jointly with an absorbed power-law model, where \texttt{tbabs} and \texttt{ztbabs} account for the X-ray absorption, and the \texttt{zdust} describes the optical extinction. The SED fitting was carried out with \texttt{HEtools} \citep{Wang2023HEtools}. The Galactic neutral hydrogen column density and Galactic reddening were fixed at $N_{\rm H,Gal}=9.31\times10^{20}\ {\rm cm^{-2}}$ and $E(B-V)_{\rm Gal}=0.0853\ \rm mag$ respectively. We adopt $9.81\times10^{20}\ {\rm cm^{-2}}$ for the foreground column in the EP/WXT and FXT reductions, and $9.31\times10^{20}\ {\rm cm^{-2}}$ for the Swift products and our uniform optical-to-X-ray SED fits. These mission-specific values differ by about 5\%, which does not imply variable absorption. The remaining host-frame absorption, host-galaxy extinction and the synchrotron spectral index were left free in the fit. We fitted these parameters separately for the prompt-emission and afterglow epochs. The two estimates are consistent within approximately \(1.8\sigma\); therefore, we adopted their midpoint, $E(B-V)_{\rm host}=0.09^{+0.02}_{-0.02}\ \rm mag$, as a representative common value in the subsequent extinction. This host-extinction value was then applied to the optical/NIR data using the \texttt{dust\_extinction} package \citep{Gordon2024} with the \texttt{G23} \citep{Gordon2023} and \texttt{G24\_SMCAvg} \citep{GordonFitzpatrick2024} extinction curves for the Milky Way and host-galaxy components, respectively. The latter was evaluated in the GRB rest-frame at $z=3.171$. We adopted a Galactic extinction ratio $R_{V,\rm MW}=3.08$ and a host-galaxy extinction ratio $R_{V,\rm host}=2.93$ (corresponding to an SMC-like extinction curve, as commonly found for GRB host galaxies; \citet{Schady2012}), giving the V-band extinctions as
\begin{equation}
A_{V}=R_{V}\,E(B-V). \label{eq:Av}
\end{equation}
Here, $A_V$ denotes the extinction in the reference optical $V$ band, centered near $0.55\,\mu\mathrm{m}$, and sets the normalization of the extinction curve. For the host-galaxy component, $A_V$ is defined in the host rest frame. The extinction in each observed optical/NIR band is evaluated at its effective wavelength $\lambda_{\rm eff}$ for the Galactic component and at $\lambda_{\rm eff}/(1+z)$ for the host component. The observed flux densities were corrected according to
\begin{align}
A_{\lambda}&=A_{\lambda,\rm Gal}+A_{\lambda,\rm host}, \label{eq:Alambda}\\
F_{\nu,\rm corr}&=F_{\nu,\rm obs}\,10^{0.4A_{\lambda}}. \label{eq:corr}
\end{align}

\begin{figure}[t!]
\centering
\includegraphics[width=\columnwidth]{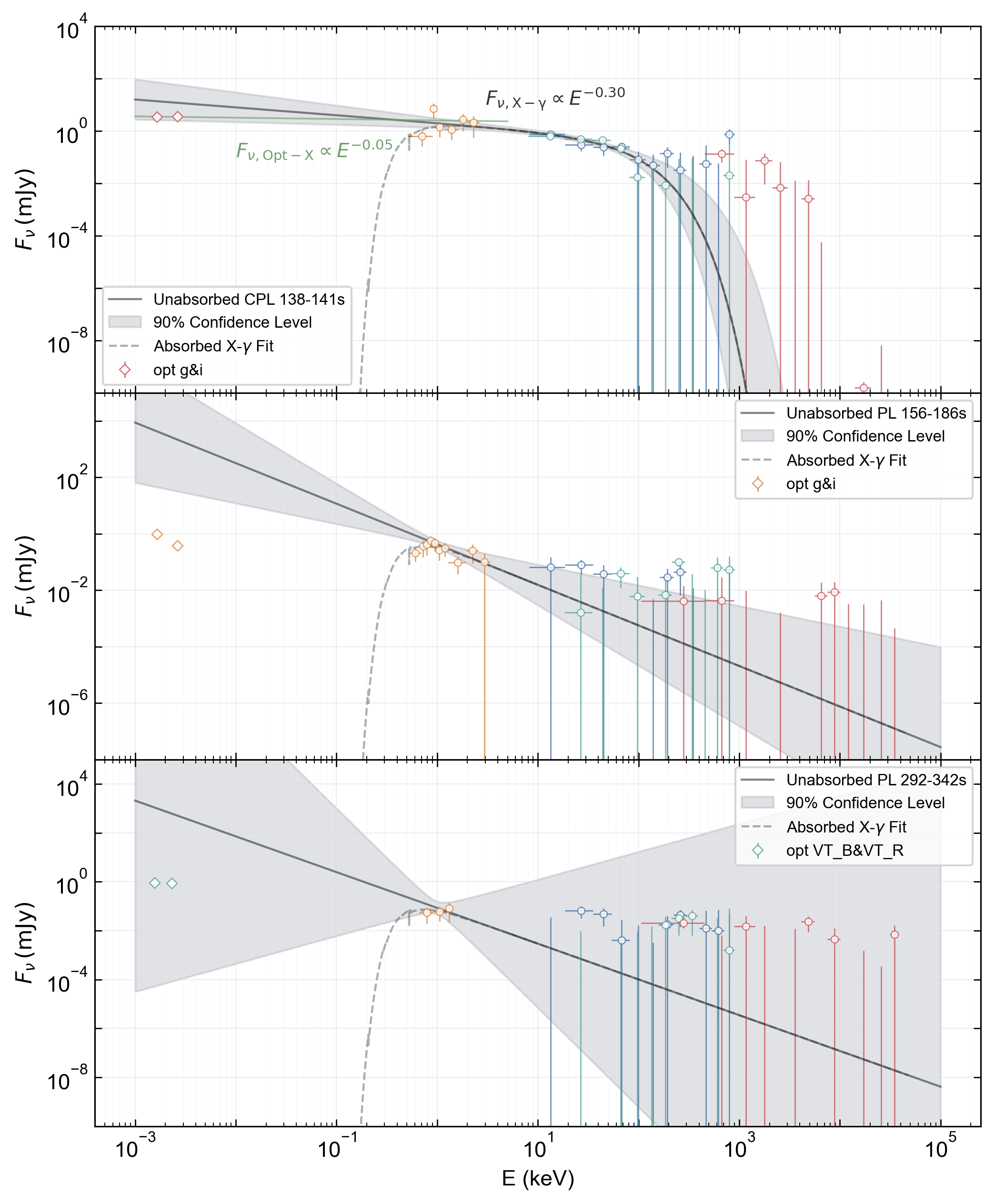}
\caption{Comparison of the extinction-corrected optical measurements with the extrapolated high-energy SED in three intervals: $T_{0,\rm EP}+[138,141]$, $[156,186]$, and $[292,342]$ s. For each interval, the \ep/WXT and \fermi/GBM spectra were fitted first, without including the optical measurements, and the resulting model was then extrapolated to the optical band. The dashed and solid black curves show the absorbed and unabsorbed models, respectively, and the shaded regions show approximate pointwise 90\% confidence envelopes calculated from the full fit covariance matrix under a local Gaussian approximation. The optical measurements are corrected for Galactic and host-galaxy extinction. The labels ``opt g\&i'' in the upper and middle panels denote the measurements obtained with the Burst Simultaneous Three-Channel Imager (BSTI) on \textit{NUTTelA-TAO} \citep{GCN43237}. The label ``opt VT\_B\&VT\_R'' in the lower panel denotes measurements in the blue (400--650~nm) and red (650--1000~nm) channels of the \svom\ Visible Telescope (VT), respectively \citep{GCN43203}. Details of the optical observations are given in Section~\ref{sec:agopt}.
\label{fig:bbsed}}
\end{figure}

The earliest optical detections overlap the prompt X-ray and gamma-ray emission. We selected three intervals centered on these points and fitted the high-energy data before extrapolating the unabsorbed model into the optical band. These intervals are filled with colors in Figure~\ref{fig:xonlc}. After extracting the \ep/WXT and \fermi/GBM spectra in these three intervals, we first fit these high energy spectra with a Band, CPL or PL model. We choose the one with the lowest BIC and then extrapolate the best fit unabsorbed model to the optical band where the early optical points are to check if they fall in the 90\% confidence region of the model. Figure~\ref{fig:bbsed} shows the corresponding spectral energy distributions (SEDs) where the optical points are corrected for the host-galaxy and Galactic extinction. The first interval provides the strongest indication that the optical and high-energy fluxes may belong to the same spectral component. We caution, however, that during the prompt phase the optical emission may contain contributions from multiple radiation components (internal shocks, reverse shock, and early forward shock), and the simple extrapolation of the high-energy spectrum into the optical band may not capture this complexity. The second interval's extrapolation envelope is higher than the optical measurments. The last interval has larger extrapolation uncertainties and do not uniquely constrain the optical region.

\section{Afterglow Modeling}\label{sec:model}

\subsection{Dynamical picture}

\begin{figure}[t]
\centering
\includegraphics[width=\columnwidth]{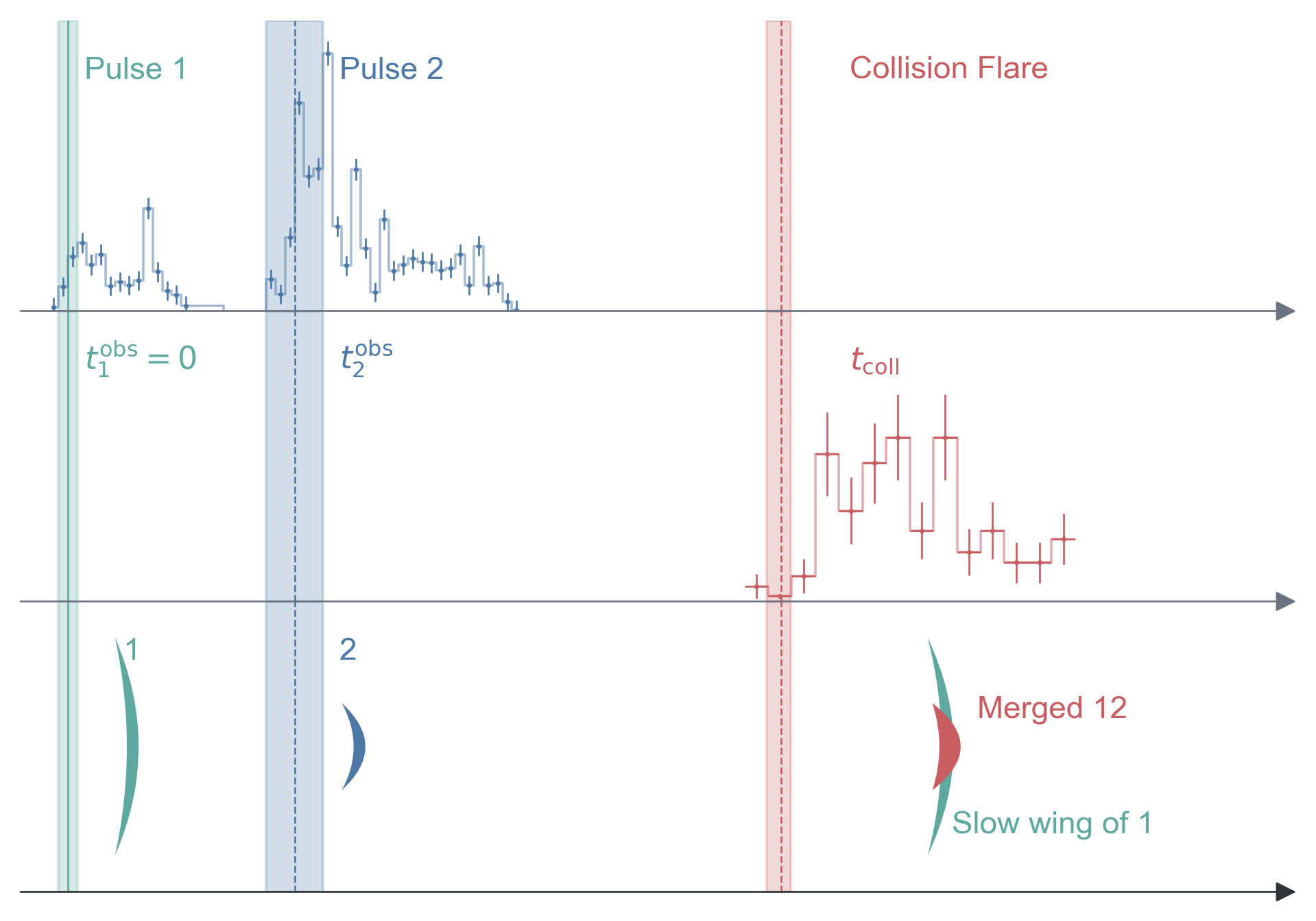}
\caption{Timing schematic of the multiple-shell collision scenario. The green solid line marks the first \fermi/GBM pulse and defines the model time origin $T_{0,\rm GBM}$, whereas the light-curve figures use the observational origin $T_{0,\rm EP}$. The purple interval shows the adopted launch-time range of the faster shell. The red interval marks the rise of the late X-ray flare, which we associate with the collision time. The ``core'' and ``wing'' are effective post-collision components rather than a hydrodynamically derived angular structure.
\label{fig:timing}}
\end{figure}

We consider a two-ejecta interpretation of the prompt pulse sequence. 
We associate the earlier gamma-ray pulse group with dissipation in a slower, wider outflow and the later, harder pulse with a faster, narrower shell. This association is not unique, because internal velocity inhomogeneity can produce several prompt pulses within one shell \citep{ReesMeszaros1994,Kobayashi1997,SariPiran1997}. 
If the later shell is faster, it can catch up with the earlier shell. Meanwhile, if the two shells have different opening angles, the collision may result in a structured jet. We assume that this collision powers the soft X-ray flare whose rise begins near $T_{0}+210$ s ($T_{0,\rm GBM}+151$ s) and whose maximum occurs near $T_{0}+220$ s. After the collision, we represent the merged inner core and the unshocked outer part of the first shell as a core and an annular wing with different Lorentz factors and energies. Figure~\ref{fig:timing} illustrates this effective geometry.

Several assumptions underlying this picture should be made explicit. 
First, we assume that the multi-pulse prompt emission implies at least two physically distinct ejecta shells, although internal velocity structure within a single shell can also produce multiple pulses. 
Second, we assume a completely inelastic collision. Equation~(\ref{eq:conservation}) follows from energy--momentum conservation in the ultrarelativistic limit, with radiative leakage during the merger small enough relative to the total shell energy such that the effects to the dynamical evolution can be neglected \citep{Geng2025}. The relative bulk kinetic energy can then be converted into internal energy. 
Third, we use the additional cold-shell approximation only to calculate the bulk kinetic energy retained by the merged ejecta from the summed baryonic rest mass, $m_{1,c}+m_2$. 
Fourth, we do not require the first shell to be angularly uniform, nor do we assume that every X-ray feature is produced by the same mechanism. In particular, the weak soft X-ray episode preceding the main gamma-ray activity is not included in the collision calculation and may instead trace jet breakout or an earlier central-engine episode. 
Fifth, we assume that the two shells are coaxial. The core--wing geometry is an effective one-zone parameterization rather than a result of a multidimensional hydrodynamic calculation. And we consider only an on-axis viewing geometry in our shell-collision model, motivated by the consistency of GRB~251222A's position in the Amati plane with the simulated on-axis GRB population \citep{GaoGeng2025}.

\subsection{Timing constraints}

For a simple chase-and-collide scenario, we can use the timing of the peaks in the light curves and the basic dynamic equations to constrain the initial Lorentz factor and the total energy of these shells, which will be a great help to model the afterglow and to understand the physical origin of the prompt emission. 
Let the observer-frame launch times of the two shells be $t_1^{\rm obs}$ and $t_2^{\rm obs}$, with velocities $\beta_1c$ and $\beta_2c$. Since the adopted launch-time separation is about 48 seconds, which is much longer than the variability timescale of the prompt emission, the radii at which the collision happens are much larger than the shell forming radii \citep{ReesMeszaros1994,Golkhou2015}.
The collision time in a chase-and-catch approximation is
\begin{equation}
t_{\rm coll}\simeq t_1^{\rm obs}
+\left(1+\frac{1-\beta_2}{\beta_2-\beta_{1,c}}\right)
\left(t_2^{\rm obs}-t_1^{\rm obs}\right),
\label{eq:tcoll}
\end{equation}
where $\beta_2>\beta_{1,c}$ and $\beta_{1,c}$($\Gamma_{1,c}$) is the velocity (bulk Lorentz factor) of the core part of the first wide shell. The counterpart to this is the wing part of the first shell with a bulk Lorentz factor of $\Gamma_{1,w}$. 

Because the second shell is assumed to be narrower, the collision merges it with the core of the first shell while leaving an outer annular component. For a completely inelastic collision with relatively small radiative loss during merging, energy--momentum conservation gives the approximate merged Lorentz factor
\begin{equation}
\Gamma_c=\Gamma_{12}=\left(
\frac{\Gamma_2m_2+\Gamma_{1,c}m_{1,c}}
{m_2/\Gamma_2+m_{1,c}/\Gamma_{1,c}}\right)^{1/2},
\label{eq:conservation}
\end{equation}
where $\Gamma_{c}(=\Gamma_{12})$ and $\Gamma_{1,w}$ denote the bulk Lorentz factors of the two parts of the resulting jet respectively, and $E_{k,c}$ and $E_{k,w}$ are the corresponding isotropic kinetic energies. Throughout this work, we adopt the cold-shell definition of the isotropic-equivalent kinetic energy, $E_k=(\Gamma-1)mc^2$, where $m$ is the corresponding isotropic-equivalent rest mass. This definition is used consistently to relate the shell masses in Equation~(\ref{eq:conservation}) to their kinetic energies.

We calculate the collision variables sequentially rather than fitting them independently. For each adopted pair of $t_2^{\rm obs}$ and $t_{\rm coll}$, Equation~(\ref{eq:tcoll}) gives $\Gamma_{1,c}$ from the sampled $\Gamma_2$. The sampled $E_{k,2}$ then gives $m_2$, and Equation~(\ref{eq:conservation}) determines the positive mass ratio $r\equiv m_{1,c}/m_2$:
\begin{align}
m_2c^2 &= \frac{E_{k,2}}{\Gamma_2-1},\notag\\
r &= \frac{\Gamma_2-\Gamma_c^2/\Gamma_2}
{\Gamma_c^2/\Gamma_{1,c}-\Gamma_{1,c}},\notag\\
E_{k,1,c} &= (\Gamma_{1,c}-1)r m_2c^2,\notag\\
E_{k,c} &= (\Gamma_c-1)(1+r)m_2c^2.
\label{eq:mass_mapping}
\end{align}

We therefore derive $E_{k,1,c}$ and $E_{k,c}$ at each likelihood evaluation instead of treating them as independent parameters. We retain only solutions with $m_{1,c}>0$ and $m_2>0$, which require $\Gamma_{1,c}<\Gamma_c<\Gamma_2$ for this two-shell configuration.
For the simplest case, the first ejected shell is not structured, which means $\Gamma_{1,c}=\Gamma_{1,w}$. To allow for angular variation in the first shell, we instead require only $\Gamma_{1,w}\leq\Gamma_{1,c}$. Overall, we require $\Gamma_{1,w}\leq\Gamma_{1,c}<\Gamma_c<\Gamma_2$.

We use isotropic-equivalent masses and energies by assuming that the coaxial shells have the same solid-angle conversion factor in their overlapping cores. Equivalently, one can write the conservation equations per unit solid angle, so that the common conversion factor cancels. If the shells are misaligned or the interacting patches have different conversion factors, Equation~(\ref{eq:mass_mapping}) requires an explicit angular-overlap model.

For a homogeneous external medium, the deceleration timescale of each component scales as
\begin{align}
t_{1,c}^{\rm{dec}}\simeq \,&t_1^{\rm obs}+1.2\times10^3(1+z)
\left(\frac{E_{k,1,c}}{10^{53}\ {\rm erg}}\right)^{1/3}\notag\\
&\times\left(\frac{n}{1\ {\rm cm^{-3}}}\right)^{-1/3}
\left(\frac{\Gamma_{1,c}}{50}\right)^{-8/3}\ {\rm s},
\label{eq:tp1}\displaybreak[0]\\
t_{c}^{\rm{dec}}\simeq \,&t_{\rm coll}+1.2\times10^3(1+z)
\left(\frac{E_{k,c}}{10^{53}\ {\rm erg}}\right)^{1/3}\notag\\
&\times\left(\frac{n}{1\ {\rm cm^{-3}}}\right)^{-1/3}
\left(\frac{\Gamma_c}{50}\right)^{-8/3}\ {\rm s},
\label{eq:tpc}\displaybreak[0]\\
t_{w}^{\rm{dec}}\simeq \,&t_1^{\rm obs}+1.2\times10^3(1+z)
\left(\frac{E_{k,w}}{10^{53}\ {\rm erg}}\right)^{1/3}\notag\\
&\times\left(\frac{n}{1\ {\rm cm^{-3}}}\right)^{-1/3}
\left(\frac{\Gamma_{1,w}}{50}\right)^{-8/3}\ {\rm s},
\label{eq:tp2}
\end{align}
which indicates two jets continue coasting from their respective formation times (the collision time $t_{\rm coll}$ for the merged core, and the first shell launch time $t_1^{\rm obs}$ for the unshocked wing) until they collect enough circumburst material to start to decelerate \citep{BlandfordMckee1976,SariPiran1999,Molinari2007,Liang2010}. Consistent with the numerical light-curve calculation, we initialize the effective merged core at $t_{\rm coll}$ and neglect its finite formation radius in the analytic scaling of Equation~(\ref{eq:tpc}). 
To ensure that the leading core is still coasting when the collision occurrs, we require $t_{1,c}^{\rm dec}>t_{\rm coll}$. 
These expressions are used as dynamical guides rather than exact observed peak-time equations. The measured peak in a given band also depends on the synchrotron break frequencies, microphysical parameters, equal-arrival-time effects, and the annular geometry of the wing.

\section{Results}\label{sec:result}

We modeled the afterglow with \texttt{VegasAfterglow}\footnote{\url{https://github.com/YihanWangAstro/VegasAfterglow}}, which evolves user-defined angular distributions of energy and Lorentz factor and computes the resulting synchrotron radiation \citep{Wang2025Vegas,Zhang2018}. 
The inner component was represented as a top-hat jet and the outer component as an annular jet, so we define the energy and Lorentz factor profiles as a piecewise function, with one set of values in the core and another in the wing. We assume an on-axis observer, $\theta_{\rm obs}=0$. The core occupies $0\leq\theta<\theta_c$, while the annular wing occupies $\theta_c\leq\theta<\theta_w$. Both propagated into the same constant-density medium. The radiation is assumed to only have the synchrotron emission of the forward shock. 

The inference combines three ingredients:
\begin{enumerate}
\item A flux/flux density likelihood comparing the model with the multi-band optical and X-ray measurements, which is calculated by the $\chi^2$ statistic. The four optical measurements obtained before $t_{\rm coll}$ are treated as one-sided upper limits on the forward-shock contribution: models below the observed fluxes are allowed, whereas models exceeding them are penalized. This treatment prevents early emission that may originate from prompt or reverse-shock processes from being incorrectly attributed to the wing forward shock, which would bias the inferred wing energy and related parameters toward artificially large values.
\item A deterministic timing and mass mapping based on Equations~(\ref{eq:tcoll}) and (\ref{eq:conservation}), together with conditions defining the allowed parameter space. We adopt an interval for every timing including the shell launch time and the collision time to account for uncertainties, while the shell launch times are set as the rising phases of each pulse and the collision time is chosen as the temporal turnover point where the X-ray flare starts to rise. The time zero point is set as $t_{1}^{\rm{obs}}$, i.e. the Fermi/GBM trigger time (GBM $T_{0}$), giving the illustrative ranges $t_{2}^{\rm{obs}}\in [42,54]\,\rm{s}$ and $t_{\rm{coll}}\in [148,153]\,\rm{s}$, shown in Figure \ref{fig:timing}, which indicate the uncertainty in associating the second pulse and the flare with the adopted launch and collision time. We tested the timing dependence on a $3\times3$ grid, comprising nine fits with $t_2^{\rm obs}=(42,48,54)$ s and $t_{\rm coll}=(148,151,153)$ s. Relative to the central grid point, $(t_2^{\rm obs},t_{\rm coll})=(48,151)$ s, the posterior medians of all 14 fitted parameters shifted by less than $0.57\,\sigma$, and the medians of the derived parameters shifted by less than $0.67\,\sigma$. The fitted afterglow parameters are quite stable across the adopted timing range. In the implemented likelihood, we set $t_1^{\rm obs}=0$ at $T_{0,\rm GBM}$ and fix $t_2^{\rm obs}=48$ s and $t_{\rm coll}=151$ s. We require the timing and mass-mapping equations, $\Gamma_{1,c}<\Gamma_c<\Gamma_2$, $m_{1,c}>0$, $m_2>0$, and $t_{1,c}^{\rm dec}>t_{\rm coll}$ to hold simultaneously. We additionally require the sample-wise dissipated energy, $E_{\rm diss}=E_{k,1,c}+E_{k,2}-E_{k,c}$, to satisfy $E_{\rm diss}\geq4.939\times10^{51}$ erg, corresponding to the upper end of the adopted 0.5--4 keV flare-energy range, which is implemented as a hard feasibility condition for our collision scenario. We set the log likelihood to minus infinity if any of these conditions is not satisfied. These hard conditions exclude regions of parameter space that are incompatible with the assumed collision dynamics or energy budget, thereby reducing the volume of the retained parameter space and alleviating some of the strong parameter degeneracies. In the numerical implementation, samples outside this support are assigned $\ln\mathcal L=-\infty$.
\item A single-band term encoding the peak time of the first afterglow bump. The peak time of the first afterglow bump in the $VT_R$ band is known from the more densely observing sequence reported by \citep{GCN43203}. We converted this single-band peak time to the GBM time reference used by the model. At each MCMC iteration, the peak is located from the total modeled $VT_R$ light curve, $F_{\nu,c}+F_{\nu,w}$. Considering that the temporal sampling is not fine enough, after identifying the candidate maximum, its peak time is interpolated by fitting a parabola in log time--log flux to that point and its two adjacent grid points, and the logarithmic-time $\chi^2_{\rm peak}=[\ln(t_{p,\rm model}/t_{p,\rm obs})/0.20]^2$ is added to the likelihood.
\end{enumerate}

We denote the detected optical/NIR and follow-up X-ray data used in the afterglow fit by $\mathcal D$. The set $\mathcal U$ contains the four pre-collision optical measurements treated as upper limits on the forward-shock contribution. We do not include the EP/WXT prompt points in the afterglow flux likelihood. The detected-data and upper-limit terms implemented in the code are
\begin{align}
\ln\mathcal L_{\rm det}={}&-\frac{1}{2}\sum_{i\in\mathcal D}
\left[\left(\frac{F_{\nu,i}^{\rm obs}-F_{\nu,i}^{\rm mod}}{\sigma_i}\right)^2
+\ln\left(2\pi\sigma_i^2\right)\right],\notag\\
\ln\mathcal L_{\rm UL}={}&\sum_{j\in\mathcal U}
\ln\Phi\left(\frac{F_{\nu,j}^{\rm obs}-F_{\nu,j}^{\rm mod}}{\sigma_j}\right),
\end{align}
where $\Phi$ is the standard normal cumulative distribution function. The optical/NIR uncertainties include the 10\% systematic floor described below. For the X-ray data, we use the reported upper or lower uncertainty according to the sign of the model residual. For parameters that satisfy the timing, mass-positivity, ordering, and coasting conditions, together with the dissipated-energy condition, the total log likelihood is
\begin{equation}
\ln\mathcal L=\ln\mathcal L_{\rm det}+\ln\mathcal L_{\rm UL}
-\frac{1}{2}\chi^2_{\rm peak};
\end{equation}
otherwise $\ln\mathcal L=-\infty$. The term $\chi^2_{\rm peak}$ supplies a phenomenological timing regularizer on the modeled first afterglow bump and conditions the posterior on the adopted two-bump decomposition. We repeated the fit with logarithmic peak-time widths of $0.10$ and $0.40$, compared with the fiducial value of $0.20$. Across these tests, the posterior medians of all sampled and collision-derived parameters remain within $0.60\,\sigma$ of the fiducial results, and their 68\% credible intervals overlap. The pulse-to-shell and flare-to-collision associations likewise enter through the adopted timing inputs and the deterministic collision mapping. Overall, the posterior support is restricted jointly by the timing relations, the mass-positivity and shell-ordering requirements, the coasting condition, and the dissipated-energy condition. However, satisfying these conditions does not provide independent evidence for the associations.

\begin{figure*}[t]
\centering
\includegraphics[width=0.8\textwidth]{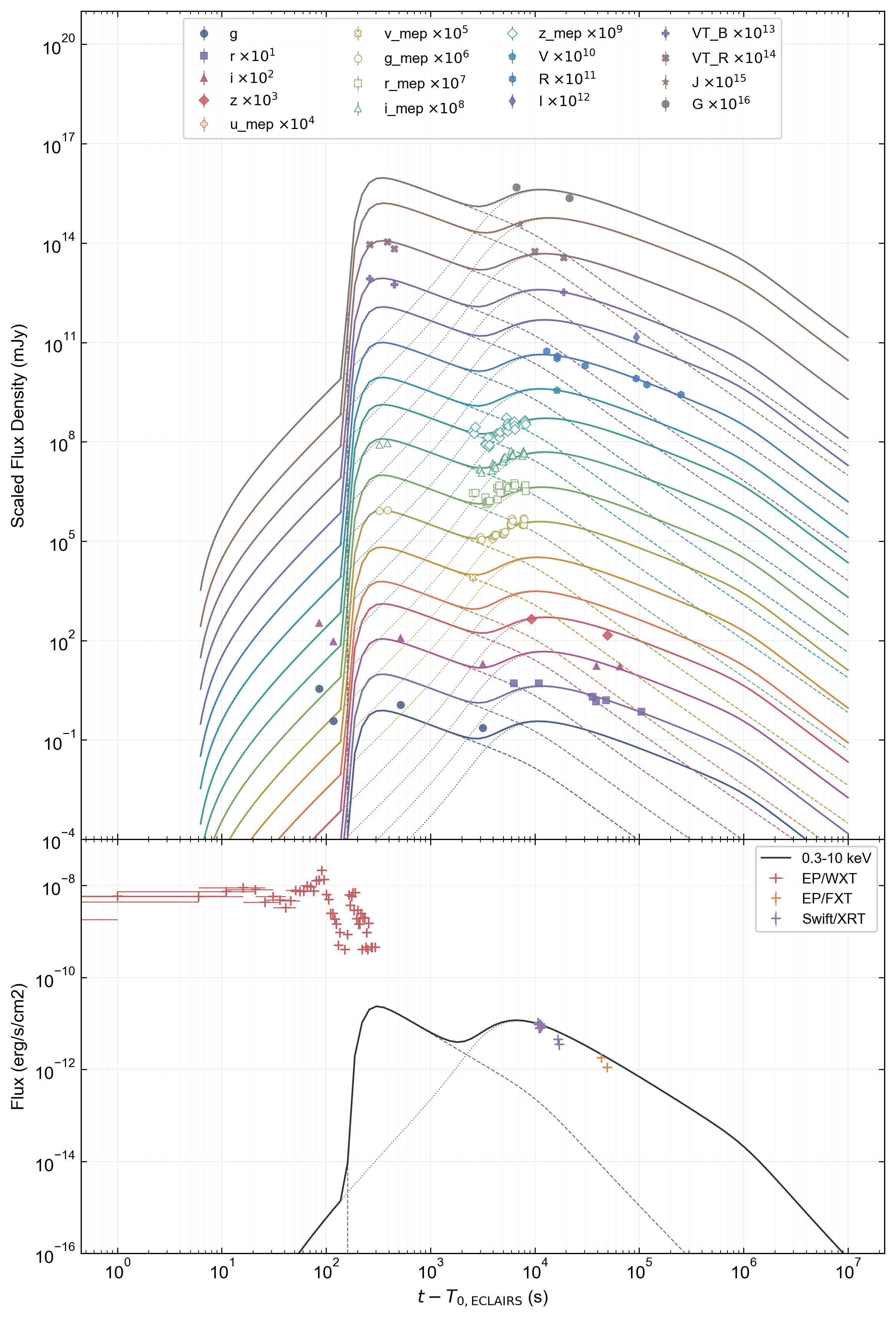}
\caption{Best fit to the optical/NIR and X-ray afterglow. Dashed and dotted curves show the inner and outer components, respectively, and solid curves show their sum. The curves correspond to the joint sample with the maximum log probability. The optical/NIR measurements and model curves are vertically rescaled by successive powers of ten, as indicated in the legend. The four optical measurements obtained before $t_{\rm coll}$ are plotted with their measured uncertainties but enter the likelihood as one-sided upper limits on the forward-shock contribution. The EP/WXT prompt data are included only for background context. The time axis is relative to $T_{0,\rm ECL}=17{:}05{:}46$ UTC, which is 5.27 s before $T_{0,\rm GBM}$ \citep{GCN43188}. We convert the collision times and the peak-time likelihood to the GBM reference defined in Section~\ref{sec:model}.
\label{fig:fit}} 
\end{figure*}

\begin{figure*}[t]
\centering
\includegraphics[width=\textwidth]{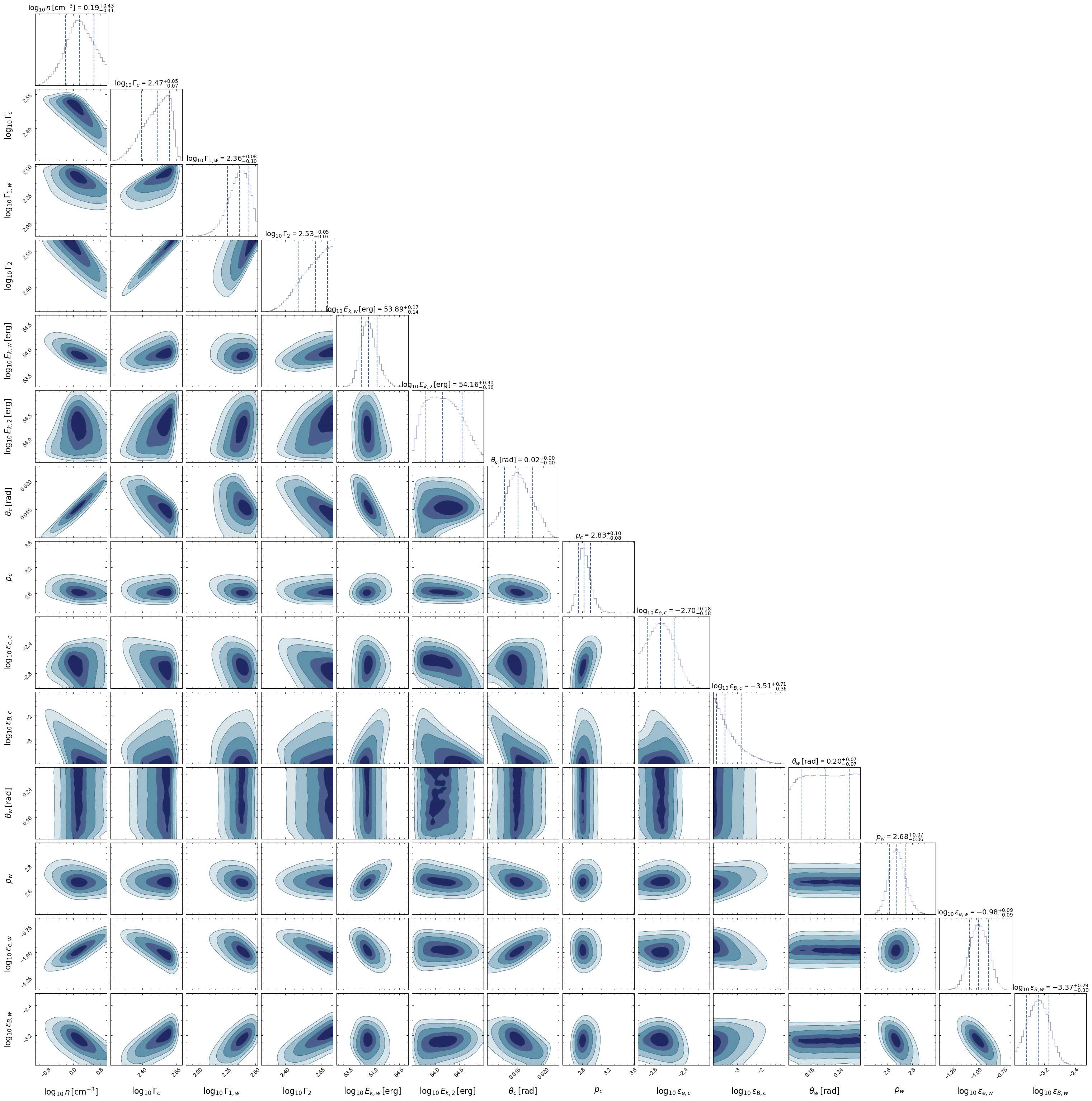}
\caption{The corner plot of the fitting result in Table~\ref{tab:param}, showing all the one- and two-dimensional projections of the posterior probability distributions of 14 parameters used in the model. The one-dimensional histograms are marginal posterior distributions of these parameters. The vertical dashed lines indicate the 16th, 50th, and 84th percentiles of the samples, respectively, which are labeled on the top of each histogram.
\label{fig:corner}}
\end{figure*}

The model has 14 free parameters: $n$, $\Gamma_c$, $\Gamma_{1,w}$, $\Gamma_2$, $E_{k,w}$, $E_{k,2}$, $\theta_c$, $p_c$, $\epsilon_{e,c}$, $\epsilon_{B,c}$, $\theta_w$, $p_w$, $\epsilon_{e,w}$, and $\epsilon_{B,w}$. We do not sample $E_{k,1,c}$ or $E_{k,c}$ independently, but recalculate both from Equation~(\ref{eq:mass_mapping}) at each likelihood evaluation. We use the Markov Chain Monte Carlo (MCMC) method to fit the observed data. The prior parameter ranges used in the MCMC fitting are summarized in Table~\ref{tab:param}. Here and throughout, $\log$ denotes the base-10 logarithm. Posterior sampling uses the \texttt{emcee} ensemble sampler \citep{emcee}. To check the reproducibility of the posterior results, we ran two independent MCMC ensembles. Both runs exceeded 100 integrated autocorrelation times for all sampled parameters and yielded consistent marginal posterior distributions. The posterior medians and equal-tailed 68\% credible intervals of the inferred parameters are also summarized in Table~\ref{tab:param}. The multiwavelength fit and the corresponding posterior projections are shown in Figures~\ref{fig:fit} and \ref{fig:corner}. These intervals depend on the adopted priors, the pulse--shell associations, and the peak-time regularizer. Figure~\ref{fig:fit} shows one maximum-posterior realization, not a posterior-predictive confidence band. We therefore do not regard its visual agreement with the data as a posterior predictive check.


\begin{deluxetable*}{ccc}
\tablewidth{0pt}
\tablecaption{Parameter Priors and Posterior Constraints.\label{tab:param}}
\tablehead{
\colhead{Parameter} & \colhead{Prior range} & \colhead{Posterior (68\%)}
}
\startdata
$\log(n~(\mathrm{cm^{-3}}))$ & $[-2.0,1.0]$ & $0.19_{-0.41}^{+0.43}$\\
$\log(\Gamma_c)$ & $[1.0,2.6]$ & $2.47_{-0.07}^{+0.05}$\\
$\log(\Gamma_{1,w})$ & $[1.0,2.6]$ & $2.36_{-0.10}^{+0.08}$\\
$\log(\Gamma_2)$\tablenotemark{a} & $[1.0,2.6]$ & $2.53_{-0.07}^{+0.05}$\\
$\log(E_{k,w}~(\mathrm{erg}))$ & $[51.0,55.0]$ & $53.89_{-0.14}^{+0.17}$\\
$\log(E_{k,2}~(\mathrm{erg}))$ & $[51.0,55.0]$ & $54.16_{-0.36}^{+0.40}$\\
$\theta_c~(\mathrm{rad})$ & $[0.01,0.06]$ & $0.02_{-0.00}^{+0.00}$\\
$p_c$ & $[2.0,4.0]$ & $2.83_{-0.08}^{+0.10}$\\
$\log(\epsilon_{e,c})$ & $[-3.0,-0.5]$ & $-2.70_{-0.18}^{+0.18}$\\
$\log(\epsilon_{B,c})$\tablenotemark{b} & $[-4.0,-1.0]$ & $-3.51_{-0.36}^{+0.71}$\\
$\theta_w~(\mathrm{rad})$\tablenotemark{c} & $[0.1,0.3]$ & $0.20_{-0.07}^{+0.07}$\\
$p_w$ & $[2.0,4.0]$ & $2.68_{-0.06}^{+0.07}$\\
$\log(\epsilon_{e,w})$ & $[-3.0,-0.5]$ & $-0.98_{-0.09}^{+0.09}$\\
$\log(\epsilon_{B,w})$ & $[-4.0,-1.0]$ & $-3.37_{-0.30}^{+0.29}$\\
\enddata
\tablecomments{$^{\rm a}$The posterior of $\Gamma_2$ remains sensitive to its upper prior boundary and is strongly correlated with $\Gamma_{1,c}$. $^{\rm b}$The $\epsilon_{B,c}$ chain is numerically converged, but its broad and asymmetric marginal posterior remains sensitive to the lower prior boundary. $^{\rm c}$The marginal posterior of $\theta_w$ remains close to the adopted prior shape and should therefore be regarded as weakly identified.}
\end{deluxetable*}

The model reproduces the broad double-peaked morphology by associating the earlier bump with the faster merged core and the later bump with the slower outer component. 
However, the sparse sampling around the first bump and the lack of continuous X-ray coverage prevent tight constraints on the wing Lorentz factor and on the relative contributions of the two components. 
With 14 free parameters and limited data, degeneracies among parameters are significant; several parameters (flagged in Table~\ref{tab:param}) are prior-limited rather than data-constrained. 
A comparison with simpler models (e.g., a single structured jet with fewer parameters) would help quantify whether the additional complexity of the two-jet model is warranted by the data. A simpler single Gaussian structured-jet implementation is in Section~\ref{sec:discussion}.

An additional caveat concerns the treatment of measurement uncertainties. The optical/NIR photometry compiled from GCN Circulars originates from multiple telescopes with different instruments, filters, and calibration procedures, and the associated systematic uncertainties are difficult to quantify consistently. Although the Mephisto data are homogeneously calibrated using Gaia XP synthetic photometry (Section~\ref{sec:agopt}), residual calibration and model-related systematics may still be present. 
To account for such uncertainties consistently, we applied a 10\% fractional systematic floor to all optical/NIR measurements, including both the GCN and Mephisto data. Specifically, the uncertainty used in the fitting was calculated as $\sigma_{\rm tot}=\sqrt{\sigma_{\rm stat}^{2}+(0.1F_{\nu})^{2}}$, where $\sigma_{\rm stat}$ is the reported statistical uncertainty and $F_{\nu}$ is the measured flux density. This treatment prevents measurements with very small formal errors from dominating the $\chi^{2}$ and can broaden the posterior distributions.

We adopt a physically motivated prior $\Gamma_{1,w}\leq\Gamma_{1,c}$, corresponding to a slower wing surrounding a faster core for the first wide shell, as angular gradients in energy and Lorentz factor are expected in GRB jets \citep{Rossi2002,ZhangMeszaros2002,Peng2005,Granot2002}. 

The pre-collision Lorentz factor $\Gamma_2$ is not directly constrained by the afterglow emission. The observed light curves primarily respond to the energy and Lorentz factor of the merged core, whereas $\Gamma_2$ enters indirectly through the collision-time relation, the mass mapping, and the derived energy budget. Moreover, in the ultrarelativistic regime the shell velocity approaches the speed of light and becomes only weakly sensitive to further increases in $\Gamma_2$. Changes in $\Gamma_2$ can consequently be compensated by changes in the shell mass, $E_{k,2}$, and the merged-core parameters while producing a similar collision time and afterglow. Physically, the data therefore require the second shell to be sufficiently faster than the first shell to produce the collision, but retain little information about its exact pre-collision Lorentz factor once the two shells have merged. The inferred $\Gamma_2$ should accordingly be regarded as upper-prior-sensitive rather than as a direct measurement from the afterglow data.

The tendency of $\epsilon_{B,c}$ toward the lower prior boundary can be understood partly through its degeneracy with the kinetic-energy scale. In the synchrotron model, the flux normalization and characteristic frequencies depend on combinations of $E_{k,c}$, $n$, $\epsilon_{e,c}$, and $\epsilon_{B,c}$ rather than on any one of these parameters independently \citep{SariPiranNarayan1998,GranotSari2002}. A lower magnetic-energy fraction can therefore be compensated by a larger kinetic energy, together with adjustments in the density and electron-energy fraction, while preserving a similar observed light curve.
Before the dissipated-energy requirement was imposed, this degeneracy also allowed a branch with comparatively larger $\epsilon_{B,c}$ and a lower overall kinetic-energy scale. Requiring the collision to provide at least the adopted upper flare energy preferentially removes these low-energy solutions and consequently suppresses the higher-$\epsilon_{B,c}$ branch. The remaining posterior is therefore shifted toward lower $\epsilon_{B,c}$ and becomes more sensitive to its lower prior boundary.

The posterior of the outer opening angle $\theta_w$ is broad and distinctly non-Gaussian. It is nearly flat across the adopted range $[0.1,0.3]$, so its median and 16th--84th percentile interval mainly reflect the center and width of this uniform prior rather than information supplied by the data. Therefore, the present fit does not provide a meaningful measurement of $\theta_w$.

We obtain $\Gamma_{c}\approx 293^{+36}_{-45}$, $\Gamma_{1,w}\approx 228^{+48}_{-48}$, $\Gamma_{2}\approx 336^{+42}_{-52}$, $E_{k,w}\approx(7.79^{+3.78}_{-2.20})\times10^{53}$ erg, and $E_{k,2}\approx(1.43^{+2.15}_{-0.81})\times10^{54}$ erg from the marginal posterior distributions (equal-tailed 68\% credible intervals). For each posterior sample, the collision-time relation gives $\Gamma_{1,c}$, and Equation~(\ref{eq:mass_mapping}) then gives the mass ratio and the two core energies. Their marginal values are $\Gamma_{1,c}\approx278^{+35}_{-43}$, $E_{k,1,c}\approx(3.7^{+3.9}_{-2.2})\times10^{54}$ erg, and $E_{k,c}\approx(5.4^{+5.4}_{-3.1})\times10^{54}$ erg. These quantities are derived for each sample and are not fitted independently. Because the mapping is nonlinear and the parameters are strongly correlated, the quoted marginal medians do not form one parameter vector and should not be substituted together into the conservation equations. The maximum-posterior sample used for the light curves has a different but internally conserving energy scale. The narrow ridges among $\Gamma_c$, $\Gamma_{1,w}$, and $\Gamma_2$ mainly result from the collision-timing relations and the coasting condition. In addition, $\Gamma_{1,w}$ is limited by $\Gamma_{1,w}\leq\Gamma_{1,c}$.

We adopt $10^{-2}\leq n\leq10\ {\rm cm}^{-3}$ in the present sampling (Table~\ref{tab:param}). This choice is physically motivated by the expectation that long GRBs occur in dense star-forming regions within their host galaxies \citep{WoosleyBloom2006}, where typical densities inferred from afterglow modeling are $n\sim 1$--$100$ cm$^{-3}$. 
In our scenario we assume that the circumburst environment is homogeneous and the same for both the core and wing jets. While we did not observe any X-ray features in afterglow and the rebrightening in the optical band can be produced by several mechanisms, the present data do not uniquely distinguish a homogeneous medium from a wind-like or density-structured environment \citep{ChevalierLi2000}. 
We also note that the absence of late-time optical and X-ray data prevents us from constraining the jet break time and hence the total kinetic energy independently of the afterglow peak time \citep{Beniamini2015}. 
The inferred density is $n\approx1.54^{+2.64}_{-0.94}\ \rm{cm}^{-3}$, which is consistent with values commonly inferred for long-GRB environments but remains degenerate with the kinetic energies and microphysical parameters \citep{WoosleyBloom2006}.

\section{Discussion}\label{sec:discussion}

Our shell collision scenario for EP251222b/GRB 251222A attributes the final core--wing structure to a radial shell collision, whereas conventional two-component models assume an angular structure at launch \citep{Rossi2002,ZhangMeszaros2002,Berger2003,Peng2005}. However, we also allow the first shell to have different core and wing Lorentz factors, and we do not calculate the angular profile hydrodynamically. Therefore, the fit does not show that the collision uniquely produces the inferred angular structure.
It also differs from the refreshed-shock scenario \citep{ReesMeszaros1998,KumarPiran2000,SariMeszaros2000,GranotNakar2003}. In our construction, the collision is required to occur before substantial external deceleration through the hard condition $t_{1,c}^{\rm dec}>t_{\rm coll}$, and it restructures the outflow, whereas a refreshed shock injects energy into an already decelerating blast wave. This coasting requirement is a model condition rather than an independent inference from the data.
We neglect lateral expansion and core--wing interaction, both treated as simple top-hat components; the real geometry is likely more complicated \citep{Rhoads1999}.

An important question is whether the multi-pulse prompt emission requires two distinct ejecta shells. Internal velocity structure within a single shell can also produce multiple pulses. 
The limited number of pulses in EP251222b does not allow us to distinguish these possibilities. The best-fitting hardening in the later gamma-ray interval and the soft late flare are qualitatively consistent with separate ejecta. 
However, their statistical significance has not been established with a dedicated model comparison. Future events with better pulse sampling may discriminate via spectral-hardness resets and bimodal lag distributions between shell groups.

The present data can also be accommodated by alternative scenarios. A pre-existing structured jet with angular gradients in energy and Lorentz factor \citep{Granot2002,Rossi2002} naturally produces multiple characteristic timescales without invoking shell collisions, and requires fewer free parameters (8 vs.\ the 14 in our model). A density enhancement in the circumburst medium \citep{NakarGranot2007} could contribute to the rebrightening if the contrast and transition scale are favorable. 
The model gives two conditional tests. First, the peak-time ratio of the two afterglow components should follow the scaling in Equations~(\ref{eq:tpc}) and (\ref{eq:tp2}). 
Second, the flare energy cannot exceed the energy dissipated in the collision. As a sensitivity test, we also examined the fit obtained without imposing this energy requirement. In the unconstrained posterior, the fraction of samples satisfying $E_{\rm diss}\geq4.939\times10^{51}$ erg is $63.4\%$. Thus, the energy-feasible region is also permitted by the data and is not created solely by the additional condition. At the same time, the unconstrained fit admits another branch with a lower collision-energy scale that cannot account for the adopted flare energy. Mathematically, the final energy-constrained posterior may be regarded as the unconstrained posterior conditioned on $E_{\rm diss}\geq4.939\times10^{51}$ erg. The condition removes this low-energy branch and narrows a region of parameter space in which the kinetic energies and microphysical parameters are strongly coupled. It finally improves the physical consistency of the retained solutions and reduces the extent of the degeneracy.
However, a flare followed by a multi-component afterglow is not unique to a shell collision. Refreshed shocks, a pre-existing angular structure, and density variations can produce similar behavior. Distinguishing these models will require a population study with the same timing and energy-budget tests, explicit alternative models, continuous X-ray coverage, and densely sampled optical afterglows.

\begin{figure*}[t]
\centering
\includegraphics[width=0.8\textwidth]{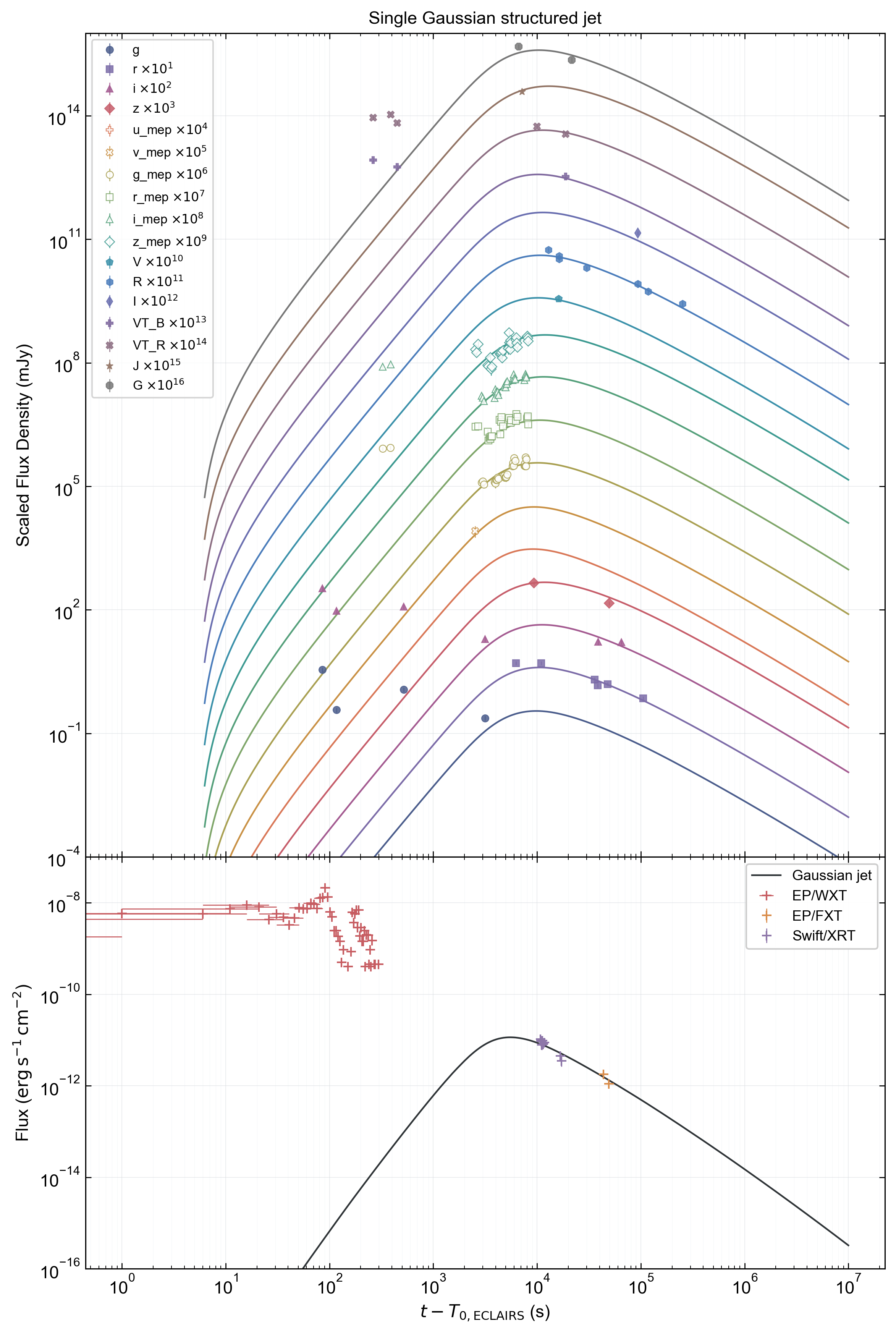}
\caption{Best-fitting multi-band afterglow light curves for the simplified single Gaussian structured-jet model. The solid curves show the predictions of the model. The EP/WXT prompt data are included only for background context. As in Figure~\ref{fig:fit}, the time axis is relative to $T_{0,\rm ECL}$, which is 5.27 s before $T_{0,\rm GBM}$. We convert the model times to the GBM reference.
\label{fig:gaussianfit}}
\end{figure*}

For comparison, we also fit the multi-band afterglow with a single structured jet. This model has eight free parameters and Gaussian angular profiles for the energy and Lorentz factor. Figure~\ref{fig:gaussianfit} shows the corresponding light curves. The resulting maximum log-likelihood is -143.57, yielding $\Delta\mathrm{BIC} = 978.73$ and $\Delta\mathrm{AIC} = 997.14$ relative to our 14-parameter model. Under the stated common data selection and likelihood definition, these information criteria provide a conditional comparison with this particular Gaussian structured-jet implementation, rather than decisive evidence for the collision scenario. A more comprehensive model comparison incorporating the prompt timing constraints within a unified Bayesian framework is deferred to future work.

Regarding the completely inelastic collision assumption (Equation~\ref{eq:conservation}), we calculate all derived energies and the corresponding efficiency separately for every joint posterior sample. The total pre-collision kinetic energy of the core region and the kinetic energy retained by the post-collision merged core are $E_{k,1,c}+E_{k,2}=\left(5.5^{+5.4}_{-3.1}\right)\times10^{54}$ erg and $E_{k,c}=\left(5.4^{+5.4}_{-3.1}\right)\times10^{54}$ erg, respectively, implying the dissipated kinetic energy is $E_{\rm diss}=\left(1.6^{+2.1}_{-0.9}\right)\times10^{52}$ erg, where the uncertainties denote the equal-tailed 68\% credible interval.
In this treatment, $E_{k,c}$ denotes the bulk kinetic energy of the cold merged shell, whereas $E_{\rm diss}$ denotes the energy converted out of bulk motion during the collision. We do not explicitly evolve the fraction of this dissipated energy that is neither emitted in the observed X-ray flare nor returned to the bulk flow.

We define the dynamical dissipation efficiency as
\begin{equation}
    \eta_{\rm diss}\equiv
    \frac{E_{\rm diss}}{E_{k,1,c}+E_{k,2}}
    =0.37^{+0.07}_{-0.14}\%.
\end{equation}
The low value is owing to the relatively modest Lorentz-factor contrast between the two shells.
We separately define the band-limited X-ray radiative efficiency as
\begin{equation}
    \eta_X\equiv
    \frac{E_{\rm flare,iso}}
    {E_{\rm flare,iso}+E_{k,c}},
\end{equation}
where $E_{k,c}$ represents the kinetic energy retained by the merged ejecta after the collision \citep{Geng2025}. 

We fit the WXT flare spectrum with an absorbed power law, using the C-statistic. The unabsorbed 0.5--4 keV flux is $F_X=(2.90^{+0.81}_{-0.64})\times10^{-9}\ {\rm erg\,cm^{-2}\,s^{-1}}$, corresponding to a 60 s fluence of $S_X=(1.74^{+0.49}_{-0.39})\times10^{-7}\ {\rm erg\,cm^{-2}}$ during $T_0+210$ to $T_0+270$ s, where the uncertainties are all the 90\% confidence limits and the fluence limits follow by multiplying the flux limits by 60 s. The resulting observed X-ray flare isotropic-equivalent energy is $E_{\rm flare,iso}\sim(3.01\text{--}4.94)\times10^{51}$ erg. With $E_{k,c}=5.4\times10^{54}$ erg and the complete posterior distribution, we obtain 
\begin{equation}
    \eta_X =
    \begin{cases}
    0.06^{+0.07}_{-0.03}\%, &
    E_{\rm flare,iso}=3.01\times10^{51}\ {\rm erg},\\[2mm]
    0.09^{+0.12}_{-0.05}\%, &
    E_{\rm flare,iso}=4.94\times10^{51}\ {\rm erg}.
    \end{cases}
\end{equation}
The posterior median of the band-limited X-ray efficiency therefore lies between approximately $0.06\%$ and $0.09\%$ across the adopted flare-energy interval. 
We emphasize that $\eta_{\rm diss}$ and $\eta_X$ characterize different aspects of the collision: the former measures the fraction of the initial bulk kinetic energy dissipated dynamically, whereas the latter compares the observed X-ray energy with the sum of the radiated and retained kinetic energies. Because $E_{\rm flare,iso}$ is evaluated in the 0.5--4 keV band, $\eta_X$ is a band-limited global X-ray efficiency. 

Among the three sets of very early optical points, the data quality is good enough to be fitted with a CPL model with a small error for the first interval. We find that its optical flux is consistent with the low-energy extension of the high-energy spectrum, as in some prompt optical GRBs \citep{Vestrand2005,Racusin2008}. 
However, this consistency alone does not show that the optical and X-ray emission arise from the same synchrotron component. Under the additional assumption that this component is synchrotron emission and that the optical band is not strongly suppressed by synchrotron self-absorption, the optical-to-X-ray spectral slope constrains the possible locations of the self-absorption, injection, and cooling frequencies ($\nu_a$, $\nu_m$, and $\nu_c$). 
For the first epoch, the low-energy slope of the high-energy spectrum shown in Figure~\ref{fig:bbsed} is $\beta_{\rm prompt}=0.30\pm0.20$, where $F_\nu\propto\nu^{-\beta_{\rm prompt}}$. The agreement of the optical points with this extrapolation supports a possible common optical-to-X-ray spectral segment with the same slope. This interpretation is based on the high-energy extrapolation, rather than a joint fit of the optical and X-ray data.
In the standard synchrotron afterglow model \citep{SariPiranNarayan1998}, $\beta=(p-1)/2$ between $\nu_m$ and $\nu_c$ in the slow-cooling regime, and $\beta=p/2$ above $\max(\nu_m,\nu_c)$. However, applying these relations requires identifying the appropriate spectral segment. We therefore do not infer a preferred cooling-break location from the present slope comparison. Other break orderings remain possible because of the unobserved frequency interval between the optical and soft X-ray bands, prompt or reverse-shock contamination, and intrinsic spectral evolution \citep{GranotSari2002}.
The second epoch's optical measurements lie below the PL extrapolation envelope, whereas the third epoch remains poorly constrained. This prevents a similar analysis for epochs 2 and 3. Rapid multicolor photometry or simultaneous infrared observations could locate a spectral turnover and test this interpretation. 

The calculation assumes a homogeneous circumburst medium, fixed microphysical parameters within each component, negligible lateral expansion, and no radiative coupling between the core and wing. These assumptions reduce degeneracy but are unlikely to hold exactly. In particular, reverse-shock emission may also contribute near an optical bump.

A population test requires more \EP\ events with prompt soft X-ray coverage and dense multi-band afterglow observations. We do not yet have a verified sample of such events. The flare and multi-component-afterglow selection functions should be defined before an association rate is estimated. We therefore use EP251222b as a case study and do not infer the occurrence rate of this behavior in the \EP\ population.

The strongest observational limitations are the X-ray gap between the prompt phase and the pointed follow-up, sparse optical sampling around the first maximum, and the absence of late measurements around a possible jet break. These gaps broaden the allowed energies, Lorentz factors, and opening angles \citep{SariPiranHalpern1999}. Future events with continuous soft X-ray monitoring and minute-cadence multiband optical observations could test whether a late prompt flare predicts a later, dynamically distinct afterglow component.

\section{Conclusions}\label{sec:conclusion}

We have presented a multiwavelength analysis of the prompt and afterglow emission of EP251222b/GRB 251222A at $z=3.171$, combining \ep/WXT soft X-ray monitoring, \fermi/GBM gamma-ray spectroscopy, \swift/XRT follow-up, and new optical/NIR photometry from Mephisto. The key observational results are:

\begin{enumerate}
\item The prompt emission contains several gamma-ray pulses and a late soft X-ray flare near $T_{0,\rm EP}+220$ s. Weak soft X-ray emission begins 42 s before the WXT onboard trigger and defines $T_{0,\rm EP}$. The time-resolved fits show a possible hardening during the later bright gamma-ray interval and a soft spectrum for the late flare. However, dedicated hypothesis tests are required to determine the significance of the hardening and the number of distinct ejecta.
\item Optical detections contemporaneous with the prompt high-energy emission show that the early optical flux is consistent with the low-energy extension of the $\gamma$/X-ray spectrum for the first epoch. This consistency alone does not identify the synchrotron spectral segment or locate the cooling break.
\item The afterglow light curve exhibits two broad optical bumps separated by roughly an order of magnitude in time, with an early maximum near several hundred seconds and a later rebrightening near $10^4$ s.
\item Host-galaxy extinction at $z=3.171$ is moderate, with $E(B-V)_{\rm host}=0.09\pm0.02\ \rm mag$ derived from joint optical-to-X-ray SED fitting assuming an SMC-like extinction curve.
\end{enumerate}

We consider a shell-collision scenario in which a faster, narrower shell catches up with the core of a slower, wider shell. Under the adopted pulse--shell and flare--collision associations, the timing relation requires $\Gamma_2>\Gamma_{1,c}$ and gives $t_{\rm coll}\approx150$ s after the GBM trigger at $T_{0,\rm GBM}$. We describe the post-collision flow by an effective merged core and an outer annular wing, rather than deriving this geometry hydrodynamically.
A joint MCMC fit to the multi-band afterglow data yields $\Gamma_c\approx 293$ and $\Gamma_{1,w}\approx 228$. The mass-conserving mapping gives $E_{k,c}\approx 5.4\times10^{54}$ erg for the merged core, while the wing energy is $E_{k,w}\approx 7.8\times10^{53}$ erg, with a circumburst density $n\sim 1.54$ cm$^{-3}$. The inferred $\Gamma_2$, $\epsilon_{B,c}$, and $\theta_w$ nevertheless retain substantial prior sensitivity or weak identifiability.

Evaluating each joint posterior sample yields $E_{\rm diss}=(1.6^{+2.1}_{-0.9})\times10^{52}$ erg and a dynamical dissipation efficiency of $\eta_{\rm diss}=0.37^{+0.07}_{-0.14}\%$. The corresponding band-limited X-ray efficiency, defined as $E_{\rm flare,iso}/(E_{\rm flare,iso}+E_{k,c})$, has a posterior median of approximately $0.06\%$--$0.09\%$ across the adopted 0.5--4 keV flare-energy interval. These two efficiencies characterize different quantities: the former measures the conversion of the pre-collision bulk kinetic energy into dissipated energy, whereas the latter compares the band-limited X-ray flare energy with the combined radiated and retained kinetic-energy budget. 
If the flare is powered by the collision and the two energies are compared using the same isotropic-equivalent convention, the observed flare represents a median fraction of approximately $18\%$--$30\%$ of the dynamically dissipated energy. These results indicate that the collision dissipates only a small fraction of the bulk kinetic energy, while a non-negligible fraction of the dissipated energy may emerge in the observed soft X-ray flare. All of these energy-budget estimates are conditional on the adopted collision dynamics requirements.

The available data are consistent with the shell-collision picture, but do not uniquely establish it. Alternative explanations -- a pre-existing structured jet with angular gradients, or density variations in the circumburst medium -- can also produce double-peaked afterglow light curves, and the current model has significant parameter degeneracies with 14 free parameters constrained by limited data. A quantitative comparison using the BIC and AIC favors the current model within the adopted fitting setup, but does not by itself provide decisive evidence against a simpler structured-jet alternative with Gaussian angular profiles for the energy and Lorentz factor.

The collision scenario provides a conditional diagnostic rather than a signature unique to this mechanism: if a late prompt-phase X-ray flare is produced by a shell collision with the core--wing geometry adopted here, it may be followed by observable multi-component afterglow evolution, provided that both components contribute appreciably to the measured flux. The characteristic peak-time ratio $t_{p,w}/t_{p,c}$ depends on the energy and Lorentz factor ratios of the two outflow components, as well as the viewing geometry, microphysical parameters, and observed spectral band. The growing sample of \EP\ GRBs with early soft X-ray coverage and multi-band optical follow-up will enable statistical tests of this timing and energy-budget relation against refreshed shocks, pre-existing jet structure, and density variations.

\appendix

\section{Optical and Near-infrared Photometry}\label{app:ondata}

\startlongtable
\begin{deluxetable*}{cccccc}
\tabletypesize{\scriptsize}
\tablewidth{0pt}
\tablecaption{Combined Optical/NIR Follow-up Observations of GRB 251222A/EP251222b. The first column lists the mid-exposure time elapsed after the adopted \ep\ $T_0$, and rows are sorted by increasing time. Magnitudes are not corrected for Galactic or host-galaxy extinction.
\label{tab:ondata}}
\tablehead{\colhead{$T-T_0$ (s)} & \colhead{Exposure (s)} & \colhead{Telescope} & \colhead{Band} & \colhead{AB Magnitude} & \colhead{Ref.}
}
\startdata
140.0 & 3 & NUTTelA-TAO/BSTI & $g$ & $17.02\pm 0.09$ & (1) \\
140.0 & 3 & NUTTelA-TAO/BSTI & $i$ & $16.20\pm 0.08$ & (1) \\
171.0 & 30 & NUTTelA-TAO/BSTI & $g$ & $19.45\pm 0.09$ & (1) \\
171.0 & 30 & NUTTelA-TAO/BSTI & $i$ & $17.58\pm 0.07$ & (1) \\
317.0 & 50 & SVOM/VT & $VT\_B$ & $18.27\pm 0.03$ & (2) \\
317.0 & 50 & SVOM/VT & $VT\_R$ & $17.57\pm 0.02$ & (2) \\
401.5 & 45 & Mephisto & $g$ & $18.31\pm 0.05$ & this paper \\
401.5 & 45 & Mephisto & $i$ & $17.60\pm 0.03$ & this paper \\
421.0 & 90 & Mephisto & $u$ & $>20.85$ & this paper \\
443.0 & 50 & SVOM/VT & $VT\_R$ & $17.37\pm 0.02$ & (2) \\
463.5 & 45 & Mephisto & $g$ & $18.26\pm 0.04$ & this paper \\
464.5 & 45 & Mephisto & $i$ & $17.49\pm 0.03$ & this paper \\
504.0 & 300 & SVOM/VT & $VT\_B$ & $18.70\pm 0.01$ & (3) \\
504.0 & 300 & SVOM/VT & $VT\_R$ & $17.89\pm 0.01$ & (3) \\
573.0 & 30 & NUTTelA-TAO/BSTI & $g$ & $18.22\pm 0.07$ & (1) \\
573.0 & 30 & NUTTelA-TAO/BSTI & $i$ & $17.32\pm 0.08$ & (1) \\
2628.0 & 60 & Mephisto & $r$ & $19.15\pm 0.08$ & this paper \\
2628.0 & 60 & Mephisto & $z$ & $18.95\pm 0.22$ & this paper \\
2686.0 & 180 & Mephisto & $v$ & $21.41\pm 0.32$ & this paper \\
2707.0 & 60 & Mephisto & $z$ & $19.14\pm 0.18$ & this paper \\
2782.0 & 60 & Mephisto & $r$ & $19.13\pm 0.08$ & this paper \\
2786.0 & 60 & Mephisto & $z$ & $18.63\pm 0.17$ & this paper \\
3008.0 & 60 & Mephisto & $g$ & $20.36\pm 0.14$ & this paper \\
3009.0 & 60 & Mephisto & $i$ & $19.38\pm 0.11$ & this paper \\
3085.0 & 60 & Mephisto & $g$ & $20.29\pm 0.14$ & this paper \\
3087.0 & 60 & Mephisto & $i$ & $19.49\pm 0.14$ & this paper \\
3162.0 & 60 & Mephisto & $g$ & $20.48\pm 0.15$ & this paper \\
3166.0 & 60 & Mephisto & $i$ & $19.71\pm 0.11$ & this paper \\
3215.0 & 180 & NUTTelA-TAO/BSTI & $g$ & $19.95\pm 0.08$ & (1) \\
3215.0 & 180 & NUTTelA-TAO/BSTI & $i$ & $19.30\pm 0.09$ & (1) \\
3430.5 & 45 & Mephisto & $r$ & $19.46\pm 0.11$ & this paper \\
3431.5 & 45 & Mephisto & $z$ & $19.90\pm 0.24$ & this paper \\
3496.5 & 45 & Mephisto & $r$ & $19.99\pm 0.11$ & this paper \\
3496.5 & 45 & Mephisto & $z$ & $19.97\pm 0.24$ & this paper \\
3541.0 & 270 & Mephisto & $v$ & $>21.67$ & this paper \\
3566.5 & 45 & Mephisto & $r$ & $19.81\pm 0.13$ & this paper \\
3635.5 & 45 & Mephisto & $r$ & $19.89\pm 0.13$ & this paper \\
3635.5 & 45 & Mephisto & $z$ & $19.43\pm 0.19$ & this paper \\
3701.5 & 45 & Mephisto & $r$ & $19.76\pm 0.10$ & this paper \\
3701.5 & 45 & Mephisto & $z$ & $20.14\pm 0.31$ & this paper \\
3766.5 & 45 & Mephisto & $r$ & $19.73\pm 0.10$ & this paper \\
3766.5 & 45 & Mephisto & $z$ & $19.99\pm 0.28$ & this paper \\
3971.5 & 45 & Mephisto & $g$ & $20.34\pm 0.15$ & this paper \\
3971.5 & 45 & Mephisto & $i$ & $19.53\pm 0.11$ & this paper \\
4038.5 & 45 & Mephisto & $g$ & $20.41\pm 0.24$ & this paper \\
4038.5 & 45 & Mephisto & $i$ & $18.98\pm 0.12$ & this paper \\
4104.5 & 45 & Mephisto & $g$ & $20.21\pm 0.14$ & this paper \\
4105.5 & 45 & Mephisto & $i$ & $19.15\pm 0.10$ & this paper \\
4171.5 & 45 & Mephisto & $g$ & $20.20\pm 0.17$ & this paper \\
4171.5 & 45 & Mephisto & $i$ & $19.23\pm 0.12$ & this paper \\
4241.5 & 45 & Mephisto & $g$ & $20.07\pm 0.13$ & this paper \\
4241.5 & 45 & Mephisto & $i$ & $19.13\pm 0.10$ & this paper \\
4307.5 & 45 & Mephisto & $g$ & $20.15\pm 0.13$ & this paper \\
4308.5 & 45 & Mephisto & $i$ & $19.33\pm 0.10$ & this paper \\
4424.5 & 45 & Mephisto & $r$ & $18.74\pm 0.08$ & this paper \\
4425.5 & 45 & Mephisto & $z$ & $19.15\pm 0.16$ & this paper \\
4493.5 & 45 & Mephisto & $r$ & $19.61\pm 0.12$ & this paper \\
4494.5 & 45 & Mephisto & $z$ & $19.25\pm 0.19$ & this paper \\
4562.5 & 45 & Mephisto & $r$ & $18.85\pm 0.09$ & this paper \\
4562.5 & 45 & Mephisto & $z$ & $19.20\pm 0.27$ & this paper \\
4629.5 & 45 & Mephisto & $r$ & $18.58\pm 0.08$ & this paper \\
4629.5 & 45 & Mephisto & $z$ & $19.12\pm 0.25$ & this paper \\
4695.5 & 45 & Mephisto & $z$ & $19.46\pm 0.22$ & this paper \\
4761.5 & 45 & Mephisto & $r$ & $19.12\pm 0.09$ & this paper \\
4761.5 & 45 & Mephisto & $z$ & $19.01\pm 0.22$ & this paper \\
4914.5 & 45 & Mephisto & $g$ & $19.95\pm 0.12$ & this paper \\
4915.5 & 45 & Mephisto & $i$ & $18.82\pm 0.09$ & this paper \\
4985.5 & 45 & Mephisto & $g$ & $20.08\pm 0.14$ & this paper \\
4985.5 & 45 & Mephisto & $i$ & $18.71\pm 0.11$ & this paper \\
5055.5 & 45 & Mephisto & $g$ & $20.08\pm 0.15$ & this paper \\
5056.5 & 45 & Mephisto & $i$ & $18.87\pm 0.14$ & this paper \\
5122.5 & 45 & Mephisto & $g$ & $20.03\pm 0.24$ & this paper \\
5122.5 & 45 & Mephisto & $i$ & $18.59\pm 0.10$ & this paper \\
5190.5 & 45 & Mephisto & $i$ & $18.65\pm 0.09$ & this paper \\
5256.5 & 45 & Mephisto & $g$ & $19.88\pm 0.13$ & this paper \\
5256.5 & 45 & Mephisto & $i$ & $18.49\pm 0.09$ & this paper \\
5372.5 & 45 & Mephisto & $z$ & $18.88\pm 0.29$ & this paper \\
5438.5 & 45 & Mephisto & $r$ & $18.62\pm 0.08$ & this paper \\
5438.5 & 45 & Mephisto & $z$ & $17.94\pm 0.14$ & this paper \\
5507.5 & 45 & Mephisto & $z$ & $18.52\pm 0.20$ & this paper \\
5575.5 & 45 & Mephisto & $z$ & $18.95\pm 0.15$ & this paper \\
5641.5 & 45 & Mephisto & $r$ & $18.93\pm 0.06$ & this paper \\
5642.5 & 45 & Mephisto & $z$ & $18.39\pm 0.15$ & this paper \\
5708.5 & 45 & Mephisto & $r$ & $18.76\pm 0.05$ & this paper \\
5709.5 & 45 & Mephisto & $z$ & $18.56\pm 0.19$ & this paper \\
5884.5 & 45 & Mephisto & $g$ & $19.38\pm 0.09$ & this paper \\
5885.5 & 45 & Mephisto & $i$ & $18.42\pm 0.06$ & this paper \\
5953.5 & 45 & Mephisto & $g$ & $19.19\pm 0.09$ & this paper \\
5954.5 & 45 & Mephisto & $i$ & $18.31\pm 0.06$ & this paper \\
6022.5 & 45 & Mephisto & $g$ & $19.36\pm 0.09$ & this paper \\
6022.5 & 45 & Mephisto & $i$ & $18.07\pm 0.06$ & this paper \\
6090.5 & 45 & Mephisto & $g$ & $18.88\pm 0.07$ & this paper \\
6091.5 & 45 & Mephisto & $i$ & $18.37\pm 0.06$ & this paper \\
6159.5 & 45 & Mephisto & $i$ & $18.32\pm 0.06$ & this paper \\
6227.5 & 45 & Mephisto & $g$ & $19.06\pm 0.07$ & this paper \\
6227.5 & 45 & Mephisto & $i$ & $18.38\pm 0.06$ & this paper \\
6344.5 & 45 & Mephisto & $r$ & $18.80\pm 0.06$ & this paper \\
6344.5 & 45 & Mephisto & $z$ & $18.36\pm 0.15$ & this paper \\
6345.0 & 300 & LCO 1 m/Sinistro (Sutherland) & $r$ & $18.55\pm 0.03$ & (4) \\
6411.5 & 45 & Mephisto & $r$ & $18.39\pm 0.04$ & this paper \\
6412.5 & 45 & Mephisto & $z$ & $18.24\pm 0.11$ & this paper \\
6477.5 & 45 & Mephisto & $z$ & $18.55\pm 0.12$ & this paper \\
6549.5 & 45 & Mephisto & $r$ & $18.56\pm 0.04$ & this paper \\
6549.5 & 45 & Mephisto & $z$ & $18.81\pm 0.15$ & this paper \\
6723.0 & 1800 & Calapai Observatory/C11 & $G$ & $18.70\pm 0.09$ & (5) \\
7254.0 & \nodata & PRIME 1.8 m & $J$ & $18.04\pm 0.02$ & (6) \\
7628.5 & 45 & Mephisto & $g$ & $19.34\pm 0.09$ & this paper \\
7629.5 & 45 & Mephisto & $i$ & $18.43\pm 0.06$ & this paper \\
7696.5 & 45 & Mephisto & $g$ & $19.21\pm 0.09$ & this paper \\
7697.5 & 45 & Mephisto & $i$ & $18.24\pm 0.06$ & this paper \\
7762.5 & 45 & Mephisto & $g$ & $19.38\pm 0.10$ & this paper \\
7762.5 & 45 & Mephisto & $i$ & $18.25\pm 0.06$ & this paper \\
7829.5 & 45 & Mephisto & $g$ & $18.85\pm 0.09$ & this paper \\
7829.5 & 45 & Mephisto & $i$ & $18.04\pm 0.07$ & this paper \\
7897.5 & 45 & Mephisto & $g$ & $19.36\pm 0.09$ & this paper \\
7897.5 & 45 & Mephisto & $i$ & $18.16\pm 0.06$ & this paper \\
7965.5 & 45 & Mephisto & $g$ & $18.96\pm 0.12$ & this paper \\
7965.5 & 45 & Mephisto & $i$ & $18.25\pm 0.06$ & this paper \\
8082.5 & 45 & Mephisto & $r$ & $18.70\pm 0.06$ & this paper \\
8082.5 & 45 & Mephisto & $z$ & $18.17\pm 0.11$ & this paper \\
8149.5 & 45 & Mephisto & $r$ & $18.62\pm 0.05$ & this paper \\
8150.5 & 45 & Mephisto & $z$ & $18.14\pm 0.10$ & this paper \\
8216.5 & 45 & Mephisto & $r$ & $18.54\pm 0.05$ & this paper \\
8216.5 & 45 & Mephisto & $z$ & $18.25\pm 0.10$ & this paper \\
8284.5 & 45 & Mephisto & $r$ & $19.02\pm 0.08$ & this paper \\
8284.5 & 45 & Mephisto & $z$ & $18.45\pm 0.18$ & this paper \\
9378.0 & 600 & LCO 1 m/Sinistro (SAAO) & $z$ & $18.17\pm 0.01$ & (7) \\
10054.0 & 100 & SVOM/VT & $VT\_R$ & $18.09\pm 0.02$ & (2) \\
10998.0 & 600 & LCO 1 m/Sinistro (SAAO) & $r$ & $18.55\pm 0.01$ & (7) \\
12973.0 & 1500 & DFOT 1.3 m & $R$ & $18.39\pm 0.01$ & (8) \\
16310.0 & 120 & ZTSh 2.6 m & $V$ & $19.14\pm 0.08$ & (9) \\
16362.0 & 900 & NOT/StanCam & $R$ & $18.96\pm 0.02$ & (10) \\
16437.0 & 120 & ZTSh 2.6 m & $R$ & $18.77\pm 0.04$ & (9) \\
18925.0 & 100 & SVOM/VT & $VT\_B$ & $19.28\pm 0.04$ & (2) \\
18925.0 & 100 & SVOM/VT & $VT\_R$ & $18.56\pm 0.03$ & (2) \\
21565.0 & 180 & GOTO-North & $G$ & $19.51\pm 0.11$ & (11) \\
30150.0 & 2700 & iTelescope T72/Kilonova-Catcher & $R$ & $19.49\pm 0.03$ & (12) \\
35874.0 & 2700 & TEC160FL/Kilonova-Catcher & $r$ & $19.55\pm 0.10$ & (12) \\
38631.0 & 15 & VLT/X-shooter acquisition image & $r$ & $19.91\pm 0.04$ & (13) \\
38826.0 & 2700 & TEC160FL/Kilonova-Catcher & $i$ & $19.45\pm 0.17$ & (12) \\
48008.0 & 4680 & SVOM/COLIBRI-DDRAGO & $r$ & $19.83\pm 0.01$ & (14) \\
49808.0 & 6180 & SVOM/COLIBRI-DDRAGO & $z$ & $19.39\pm 0.01$ & (14) \\
65048.0 & 3600 & Lulin SLT/Kinder & $i$ & $19.48\pm 0.07$ & (15) \\
93310.0 & 4320 & AbAO/AS-32 0.7 m & $R$ & $20.46\pm 0.18$ & (16) \\
93887.0 & 1600 & DFOT 1.3 m & $I$ & $19.55\pm 0.02$ & (17) \\
105302.0 & 900 & FTW 2 m/3KK & $r$ & $20.68\pm 0.02$ & (18) \\
118134.0 & 3600 & iTelescope T72/Kilonova-Catcher & $R$ & $20.92\pm 0.08$ & (12) \\
251859.0 & 3600 & Mondy/AZT-33IK 1.5 m & $R$ & $21.66\pm 0.12$ & (19) \\

\enddata
\tablecomments{References: (1) \citet{GCN43237}; (2) \citet{GCN43203}; (3) \citet{GCN43189}; (4) \citet{GCN43212}; (5) \citet{GCN43195}; (6) \citet{GCN43197}; (7) \citet{GCN43208}; (8) \citet{GCN43207}; (9) \citet{GCN43226}; (10) \citet{GCN43194}; (11) \citet{GCN43209}; (12) \citet{GCN43229}; (13) \citet{GCN43204}; (14) \citet{GCN43225}; (15) \citet{GCN43236}; (16) \citet{GCN43230}; (17) \citet{GCN43224}; (18) \citet{GCN43233}; (19) \citet{GCN43273}. \mephisto\ measurements are marked as ``this paper'', which also include the GCN reported points by \mephisto\ \citep{GCN43250}. Exposure values are total integrations when a Circular reports a stacked measurement.}
\end{deluxetable*}

\begin{acknowledgments}

    This work is partially supported by the following funding programs: the National Natural Science Foundation of China (grant Nos. 12373052, 12622309, 12273113, 12321003, 12393812, 12393813), the Strategic Priority Research Program of the Chinese Academy of Sciences (grant No. XDB0550400), the National Key R\&D Program of China (2023YFE0101200), and the Yunnan Revitalization Talent Support Program (YunLing Scholar Project).  

    This work is based on the data obtained with Einstein Probe, a space mission supported by the Strategic Priority Program on Space Science of Chinese Academy of Sciences, in collaboration with the European Space Agency, the Max-Planck-Institute for extraterrestrial Physics (Germany), and the Centre National d’Études Spatiales (France) (grant No. XDA15310000). J.-J.G. acknowledges support from the Youth Innovation Promotion Association (2023331).

    Mephisto is developed at and operated by the South-Western Institute for Astronomy Research of Yunnan University (SWIFAR-YNU), funded by the ``Yunnan University Development Plan for World-Class University" and ``Yunnan University Development Plan for World-Class Astronomy Discipline". The Mephisto team acknowledges support from the Key Laboratory of Survey Science of Yunnan Province (202449CE340002), the ``Science \& Technology Champion Project'' (202005AB160002), and from two ``Team Projects" -- the ``Top Team'' (202305AT350002) and the ``Innovation Team'' (202105AE160021), funded by the ``Yunnan Revitalization Talent Support Program". They also acknowledge two National Key Research and Development Programs of China (2024YFA1611603, 2024YFA1611604). B.K. thanks support from the ``Special Project for High-End Foreign Experts", Xingdian Funding from Yunnan Province, China. G.-W.D. also acknowledges support from the National Natural Science Foundation of China (NSFC) under grant No. 12603161.

    This work also made use of data supplied by the UK Swift Science Data Centre at the University of Leicester, and data and software obtained from the High Energy Astrophysics Science Archive Research Center (HEASARC), a service of the Astrophysics Science Division at NASA/GSFC and of the Smithsonian Astrophysical Observatory's High Energy Astrophysics Division.

\end{acknowledgments}






\facilities{\EP\ (WXT, FXT), \mephisto, \fermi\ (GBM), \swift\ (XRT)}

\software{Astropy \citep{Astropy2013,Astropy2018,Astropy2022}, XSPEC \citep{Arnaud1996}, emcee \citep{emcee}, corner \citep{corner}, VegasAfterglow \citep{Wang2025Vegas}}

\bibliography{ep251222b}
\bibliographystyle{aasjournalv7}

\end{document}